\documentclass[10pt]{IEEEtran}
\IEEEoverridecommandlockouts
\usepackage{cite}
\usepackage{textcomp}
\usepackage{xcolor}
\def\BibTeX{{\rm B\kern-.05em{\sc i\kern-.025em b}\kern-.08em
    T\kern-.1667em\lower.7ex\hbox{E}\kern-.125emX}}

\usepackage{float}
\usepackage[inline]{enumitem}
\usepackage{array}
\usepackage{tabularx}
\usepackage{booktabs}
\usepackage{makecell}
\usepackage{pifont}

\usepackage{amsmath,amssymb,amsfonts}
\usepackage[endLComment=]{algpseudocodex}
\usepackage[ruled]{algorithm}

\usepackage{comment}
\usepackage{url}

\usepackage{xspace}
\usepackage{graphicx}
\usepackage{colortbl}
\usepackage{listings}
\usepackage{placeins}
\usepackage{shuffle}
\usepackage{subcaption}
\usepackage{ifthen}
\usepackage{adjustbox}

\setlist[itemize]{nosep, leftmargin=1em}
\setlist[description]{nosep}
\setlist[enumerate]{nosep}

\lstdefinelanguage{CUDA}{
  language=C++,
  morekeywords={
    __global__, __device__, __host__, __kernel__, 
    __constant__, __shared__, threadIdx, blockIdx, 
    blockDim, gridDim, cudaMemcpy, cudaMalloc, 
    cudaSuccess, cudaMemcpyHostToDevice, cudaMemcpyDeviceToHost
  }
}

\newcommand{\regtm}{\textsuperscript{\textregistered{}}}

\begin{document}

\title{Enumeration-based Triangle Counting on GPUs}
\author{
    \IEEEauthorblockN{Sayan Ghosh,}
    \IEEEauthorblockA{Pacific Northwest National Laboratory, Email: sg0@pnnl.gov}
}
\maketitle

\thispagestyle{plain}
\pagestyle{plain}

\begin{abstract}
Subgraph enumeration in partitioned graphs across distributed-memory systems is beset with challenges in balancing data movement with the local computation. Especially, counting the number of triangles induced by the closed wedges (i.e., length-3 cycles) is a measure of connectedness in a graph, but often limited by the significant search space caused by open wedges or length-2 paths (usually greater than the number of edges in a graph), which must be inspected to assess whether a particular wedge is closed. To mitigate triangle counting on massive graphs, distributed-memory partitioned-graph approaches are viable (i.e., assessing closed wedges independently within a partition), but load imbalance due to arbitrary communication can overshadow any gains from enhanced concurrency, owing to the inherent lack of structure in regular graphs.

In this paper, we discuss a distributed-memory exact graph triangle counting implementation, and assess its performance over billion-edge real-world graphs on contemporary scale-out and scale-up based NVIDIA\regtm{} GPU platforms (considering NVIDIA ``Ampere'' and ``Hopper'' architectures), depicting a scalability spectrum of multiple orders of magnitude. Comparisons with existing state-of-the-art approaches demonstrates up to 10$\times$ improvements relative to distributed-memory CPU baseline, 2--100$\times$ over single-GPU vendor-optimized NVIDIA RAPIDS\regtm{} \textsc{cuGraph} and 10--60\% over multi-GPU $G^2$Miner, in terms of end-to-end execution times.

\end{abstract}
\begin{IEEEkeywords}
Graph Analytics, Distributed-memory, MPI, Triangle counting, GPU
\end{IEEEkeywords}

\section{Introduction}\label{sec:intro}

While a number of complex graph algorithms can be practically solved using approximation algorithms and heuristics (which trade-off computation costs with accuracy), the need for exact algorithms remains for fundamental discrete computations that shed light on the overall structure and stability of the graph or network. One of the simplest computation is calculating the number of triangles or paths of length-3 in a graph. The number of triangles augment several algorithms that quantitatively expresses the connectivity within a graph~\cite{al2018triangle}. However, enumerating triangles at scale requires careful balancing of local computation with associated long-latency data movements, due to prevalence of distance-2 computations (i.e., involving neighbor of a neighbor), causing significant load imbalances. On the contrary, graph traversals such as Breadth-First Search (BFS) exhibit distance-1 computations (i.e., only involving a current neighbor) and fixed-size frontiers, keeping the intermediate memory and associated computations directly proportional to the size of the distributed graph. 

For large graphs, the amount of concurrency can be increased via distributed-memory (within and across compute nodes), requiring careful reconciliation of the available concurrency with rising data movement and locality. Contemporary GPUs can indeed enhance the throughput of processing local subgraphs, but can worsen the load imbalance due to expanding the gap between processing and data access. Optimal GPU concurrency requires maximizing the \#thread blocks that can be executed in parallel and assumes uniform execution duration of the thread blocks, often challenging to realize for partitioned-graph triangle counting workloads due to non-existent data reuse potential and difficulty in mapping general graphs on multidimensional GPU grid-blocks more suitable for dense or structured-sparse data, causing unavoidable stalls in the data pipeline.  

\begin{figure}[!ht]
\centering
\includegraphics[width=0.49\textwidth]{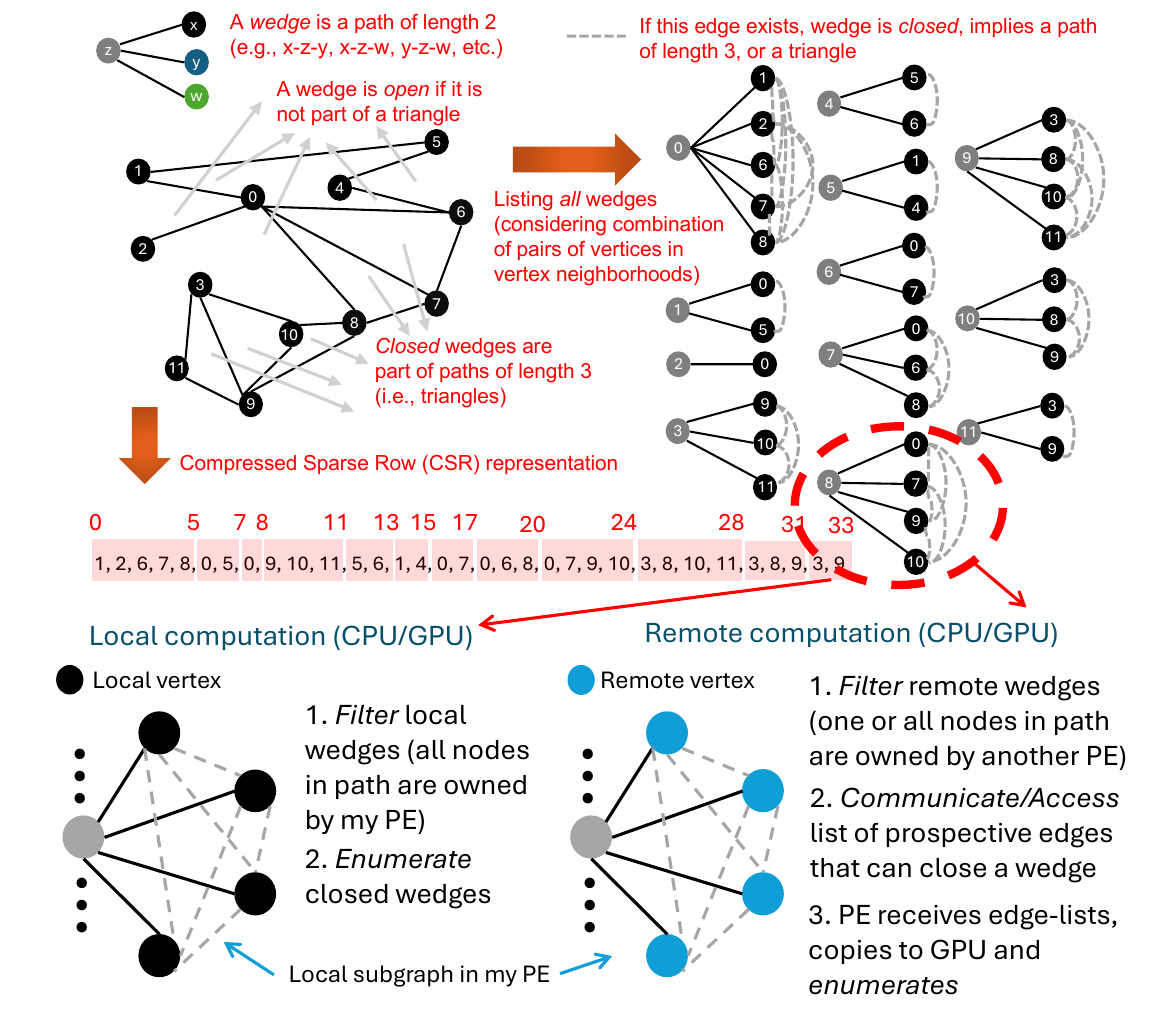}
\caption{\small The process of counting the triangles (i.e., 3-hop paths) in a graph require \emph{filtering} out wedges (i.e., 2-hop paths) and subsequently \emph{enumerating} if they are closed. Associated local\slash global tasks for a single subgraph for a partitioned graph is shown.}
\label{fig:triangles-cartoon} 
\end{figure}

Fig.~\ref{fig:triangles-cartoon} provides an intuitive representation of triangles induced by 3-length paths in an arbitrary graph, in terms of wedges (length-2 paths), and associated local and distributed-memory computations. Graphs are usually represented in Compressed Sparse Row (CSR) format to store the list of edges (in the form of arrays of indices and edges), and consecutive pairs of edges in CSR can be considered as wedges, considering sorted edge-list. A wedge is ``open'' if it is not part of a triangle (i.e., 3-length paths), whereas it is considered to be ``closed'' otherwise. To mitigate the substantial amount of concurrent distance-2 tasks in triangle counting (proportional to the number of open wedges), multithreaded CPUs\slash GPUs can be used, in conjunction with distributed-memory. 
Currently, prominent methodologies for triangle counting are based on subgraph matching, set intersection and sparse-matrix multiplication~\cite{strausz2022asynchronous,steil2021towards,pearce2019one,wang2016comparative}. Broadly, these methods have two distinct phases: \emph{filter} and \emph{enumerate}. In the \emph{filter} phase, vertices not contributing to the final solution are pruned (e.g., $LU$ matrix decomposition, advance-filter, etc.), optionally building a hash table out of the candidate edges for faster lookup, avoiding neighborhood scans~\cite{chen2022efficient, pandey2021trust, chen2021sandslash, bisson2017high}. Subsequently, the \emph{enumerate} phase identifies triangles among batches of the filtered candidate edges through set intersections (often considering degree order for reducing redundant computations, e.g., forward methodology~\cite{wang2017gunrock} and row-permutations~\cite{azad2015parallel}) among respective neighbor edge-lists (if $\{u,v\}$ is a candidate edge, then the number of triangles formed is $\mathcal{N}(u) \cap \mathcal{N}(v)$), where $\mathcal{N}(x)$ represents neighbor edge-list of an arbitrary vertex $x$) or element-wise product between permuted graph adjacency matrix and the corresponding LU factors (if $A$ is the adjacency matrix and $\hat{A}$ represents degree-based row permutation, $\hat{A} \odot (LU)$ operation obtains the exact number of triangles, counted twice). 

TriC~\cite{ghosh2020tric, ghosh2022improved} provides another perspective to triangle counting by exploiting CSR --- performing a search per wedge to determine if it is closed. If a graph has $|V|$ vertices $|E|$ edges (usually $|E| >> |V|$), this scheme would require a maximum of ${}^{|V|} C_2$ enumerations, but with filtering, substantial number of edge lists are eliminated. Our approach is derived from this idea, fusing aspects of the \emph{filter} and \emph{enumerate} phases applied to batches of edge-lists for devising GPU-based triangle counting. Since modern GPUs have hundreds of thousands of concurrent active threads, each thread can perform the requisite enumerations in parallel, providing better parallel efficiency than multithreaded CPU implementations since relative \#enumerations\slash GPU thread is less than \#enumerations\slash CPU thread (considering similar load imbalance and communication overheads). 

Further, existing shared and distributed-memory exact triangle\slash motif counting approaches considers conditional computing (to facilitate ``single'' counting), edge replication (i.e., communication free), message aggregation\slash buffering, serialization, reordering and transparent caching to improve the performance at scale~\cite{chen2022efficient, pandey2021trust, chen2021sandslash, pearce2019one, ghosh2022improved, bisson2017high, besta2022probgraph, strausz2022asynchronous}. This work adopts some of the above strategies in devising GPU-enabled distributed-memory triangle counting, without making special adaptations to the core graph data structure or partitioning (i.e., no intermediate hash tables, probabilistic representations or node\slash edge replication in distributed context), performing extensive empirical evaluations and analysis. We empirically show that the proposed approach is performant across scaling domains and diverse inputs.

Our contributions are:
\begin{enumerate}[left=0pt, nosep]
    \item Propose adaptations for single exact counting of triangles in undirected graphs on distributed-memory GPU-based systems, devising asynchronous buffered baseline CPU implementation using MPI nonblocking interface, adapting to GPU variants (\S\ref{sec:method}).
    \item Detailed performance assessments (baseline CPU\slash GPU performance, zero-copy IPC, hybrid CPU-GPU, distributed-memory GPU variants, etc.) on NERSC Perlmutter, ALCF Polaris and NVIDIA\regtm{} DGX\regtm{} dense-GPU systems (with NVIDIA\regtm{} A100\regtm{} and H100\regtm{} GPUs) using diverse input graphs (\S\ref{sec:eval}).
    \item Extensive performance comparisons with state-of-art graph frameworks and implementations, considering distributed-memory CPU (\textsc{TriC}~\cite{ghosh2020tric}, \S\ref{sec:eval-mpi}) and single\slash multi-GPU (NVIDIA Rapids \textsc{cuGraph}~\cite{fender2022rapids}, TRUST~\cite{pandey2021trust} and $G^2$Miner~\cite{chen2022efficient}, \S\ref{sec:related}).
\end{enumerate}

\section{Distributed-Memory Methodology}\label{sec:method}
Graph partitioning is NP-hard;~\cite{bui1992finding} we consider a vertex-based partitioning strategy that balances the  \#edges\slash Processing Element (PE) to minimize the standard deviation of the edge distribution across PEs\footnote{We use PE\slash process\slash rank to indicate an MPI process.}. 
\begin{figure*}[!ht]
\centering
\includegraphics[width=\textwidth]{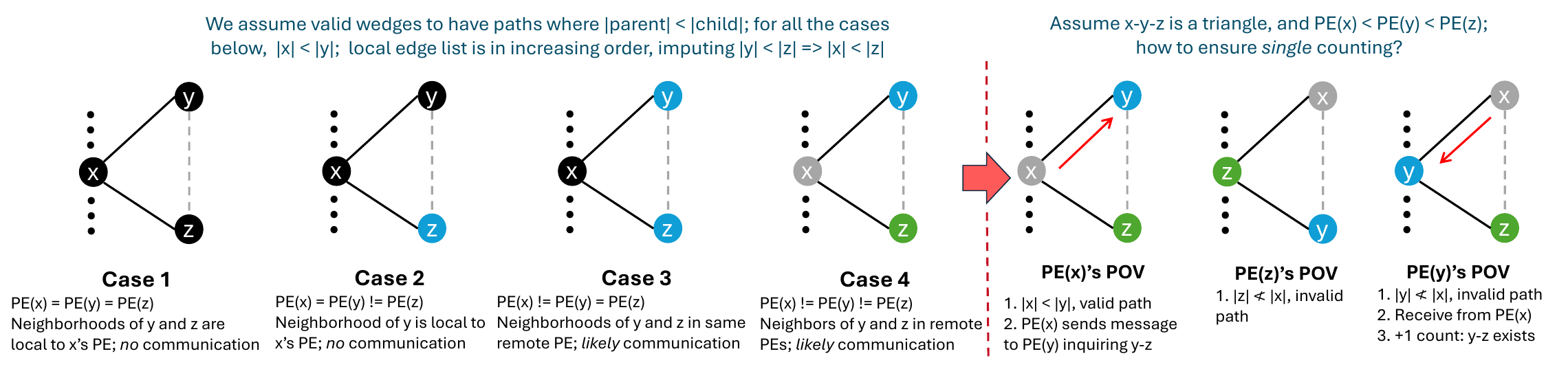}
\caption{\small Distributed counting precedence rules to enable single counting by ignoring invalid paths. Vertices with the same color are owned by the same PE. Case 4 depict scenario where each vertex is owned by a distinct PE, demonstrating single counting across PEs.}
\label{fig:counting-precedence} 
\end{figure*}
The graph is initially partitioned on CPUs (uses MPI collective I/O routines to read the binary graph data), the partitions or subgraphs (represented as CSR) are copied into distinct GPUs before initiating the local counting tasks. 
\begin{figure}[!ht]
\centering
\includegraphics[width=0.49\textwidth]{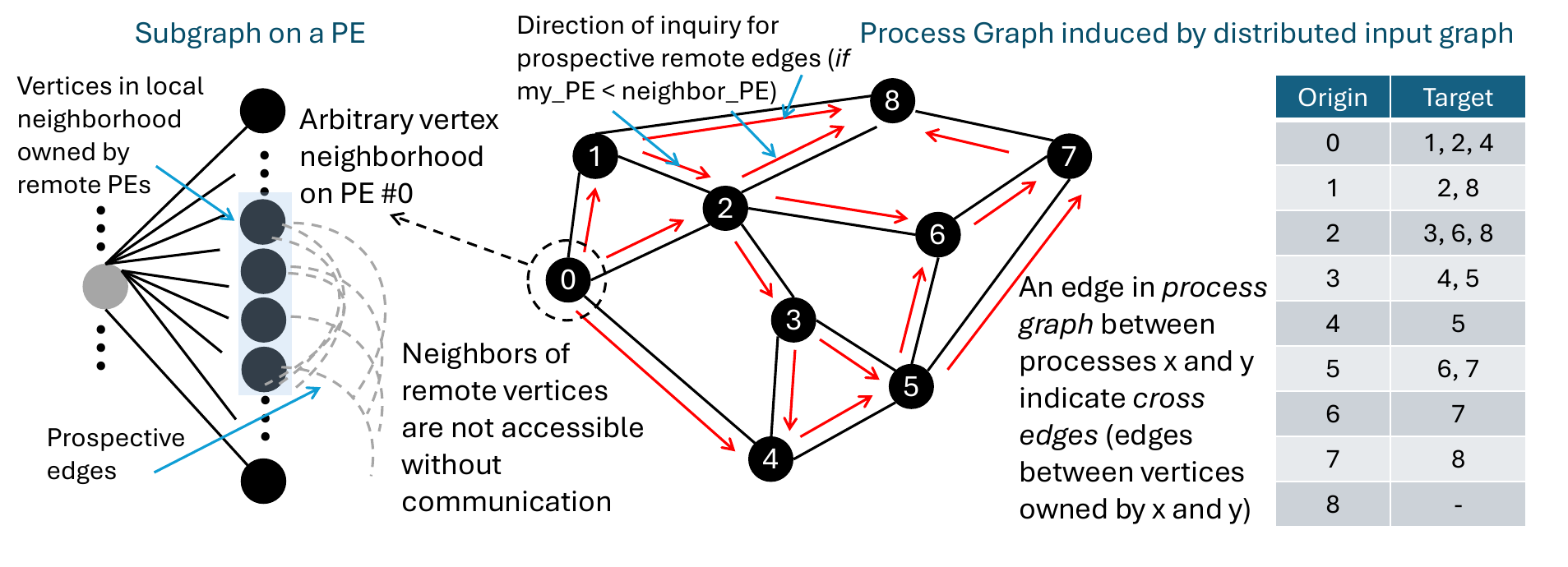}
\caption{\small Enabling directed remote inquiry for single counting (i.e., $|parent| < |child|$).}
\label{fig:remote-inquiry} 
\end{figure}
Partitioned undirected graphs must replicate edges across PEs, creating duplicate work or specific edge-access order requirements during independent triangle enumerations. For instance, matrix multiplication based triangle counting implementations~\cite{wang2016comparative,bari2021a64fx} requires division by $2$, whereas intersection-based approaches ensure symmetry order for enumerating subgraphs once~\cite{bisson2017high,chen2021sandslash}. But, single counting requires arbitrary control flows, causing warp\footnote{groups of threads executing tasks in a lock step} branch divergence (threads within a warp taking different paths) and serialization on GPUs: we perform host-based intermediate filtering of edges to avoid such issues on device. To enable single counting (and minimizing communication), we define a \emph{valid path} between an arbitrary \emph{parent} and \emph{child} vertex must require $|parent| < |child|$: consequently a PE (parent) can send messages to its neighbors (children) as long as $|PE| < |neighbor\; PE|$. We present MPI-based distributed-memory implementation in \S\ref{ssec:method-mpi-dist}, and \S\ref{ssec:method-gpu-tric} introduces the GPU methods.

We present the possible cases considering a triangle formed by three arbitrary vertices in a graph: $\widehat{x-y-z}$, establishing rules to determine if a path is valid to enable single counting. In Fig.~\ref{fig:counting-precedence}, vertices $y$ and $z$ are part of vertex $x$'s neighborhood. The problem of assessing whether $\widehat{x-y-z}$ is a triangle is then about determining whether edge $y-z$ exists. We decide whether the path from an arbitrary parent to child vertex is valid (vertices in the neighborhood of a particular vertex can be considered as children of the parent vertex). In Fig.~\ref{fig:counting-precedence}, for the wedge $x-\{y|z\}$ to be valid ($x$ is the parent vertex and $y$\slash $z$ are children), following inequalities must hold: $|x| < |y| \; \mathit{and}\; |y| < |z|  \implies |x| < |z|$. It is apparent that the directed inquiry rules for the process graph as shown in Fig.~\ref{fig:remote-inquiry} also adhere to the established precedence rules, i.e., any PE (parent) can send messages to its neighbors (children) as long as $|PE| < |neighbor\; PE|$. In Fig.~\ref{fig:counting-precedence}, we expand Case 4 to demonstrate that single counting precedence rules hold even when each vertex in the wedge $\widehat{x-y?z}$ belongs to a different PE.

\subsection{Distributed-memory implementation}\label{ssec:method-mpi-dist}
We adopt an iterative method as shown in Algorithm~\ref{alg:tric-algorithm} which continues until all incoming\slash outgoing messages to enumerate the prospective edges that can close existing wedges are processed. Distinctions between the CPU and GPU versions are discussed in \S\ref{sssec:alg-cpugpu}, followed by GPU methodologies in \S\ref{ssec:method-gpu-tric}.
\subsubsection{Local counting and posting messages}\label{sssec:alg-localcount}
Lines~\ref{alg:line:local-counting-post-begin}--\ref{alg:line:local-counting-post-end} is about local counting  and preparing outgoing user-defined fixed buffers (i.e., $SIZE$) for remote edge lookup. As shown in Lines~\ref{alg:line:local-counting-begin}--\ref{alg:line:local-counting-end}, an arbitrary wedge $\widehat{v-m?n}$ is closed (i.e., $\widehat{v-m-n}$ is a triangle) if edge $\{m, n\}$ exists (in the neighborhood of vertex $m$). If the outgoing buffer marked for a particular process is full, we initiate communication and resume scanning the local subgraph (for e.g., see Lines ~\ref{alg:line:post-outgoing-begin}--\ref{alg:line:post-outgoing-end}). We omit some details for simplicity: such as keeping track of the suspended subgraph ranges (in case a buffer is full) to continue the work exactly from where it had been left (to avoid spuriously iterating over the entire local subgraph) and potential optimizations to reduce data transfer overheads (e.g., compressing communication buffers). The status of the outgoing sends are checked at every super-step (Lines~\ref{alg:line:test-posted-isends}--\ref{alg:line:local-counting-post-end}), and buffers\slash counters\slash MPI request handles pertaining to the completed send operations are reset for reuse. The final round of pending messages are flushed out as the outstanding message count becomes zero, indicating (local) completion of the task of enumerating relevant remote edges (Lines ~\ref{alg:line:flush-sends-begin}--\ref{alg:line:flush-sends-end}). 

\begin{algorithm}[!ht]
{\scriptsize
\caption{ 
\small
Buffered graph triangle counting using asynchronous MPI nonblocking peer-to-peer interface.
\newline \textbf{Input}: $G_i=(V_i,E_i)$ portion of the (undirected) graph $G$ on Process $i$, containing ``local'' and ``ghost'' vertices (not owned by process $i$) that have an edge with a vertex $v \in V_i$. $SIZE$ is the outgoing\slash incoming buffer size.
\newline \textbf{Output}: Number of triangles $\tau$ in $G$.}
\label{alg:tric-algorithm}
\begin{algorithmic}[1]
  \State \Comment{{\bf \textcolor{blue}{--- Initialization ---}}}
  \State $\tau \leftarrow 0$ \Comment{number of triangles}
  \State $\{bar, done\} \leftarrow false, nb\_bar \leftarrow$ \texttt{MPI\_COMM\_NULL} \Comment{track completion}
  \State $sources, targets \leftarrow$ \{ \emph{list} of neighboring processes to \emph{send to} \slash \emph{receive from} \}
  \State $rbuf[0..|sources|\times SIZE] | sbuf[0..|targets|\times SIZE] \leftarrow \Phi$ \Comment{buffers}
  \State $rreqs[0..|sources|] \leftarrow \Phi,sreqs[0..|targets|] \leftarrow \Phi$ \Comment{MPI request handles}
  \State $recv\_count[0..|sources|] \leftarrow$ \{ count of elements to \emph{receive} from\} \Comment{counters}
  \State $incoming\_msgs \leftarrow  \sum(recv\_count[0..|sources|])$
  \State $outgoing\_msgs \leftarrow \sum(send\_count[0..|targets|])$
  \State $\forall {x, t \in \mathtt{enumerate}(targets)} \Rightarrow msg\_sent[x] = false$ \Comment{outgoing statuses}
  \For{$y, s \in \mathtt{enumerate}(sources) \mid recv\_count[y] > 0$} \Comment{prepost incoming} \label{alg:line:prepost-recvs}
  \State $\mathtt{MPI\_Irecv}(rbuf[y], SIZE, s, rreqs[y])$
  \EndFor
  \While{!done} \Comment{Super-step} \label{alg:line:superstep-begin}
  \State \Comment{{\bf \textcolor{blue}{--- Perform local counting and post outgoing messages ---}}}
  \If{$outgoing\_msgs == 0$} \Comment{flush outstanding messages} \label{alg:line:flush-sends-begin}
  \For{$x, t \in \mathtt{enumerate}(targets) \mid \; !msg\_sent[x]$}
  \State $\mathtt{MPI\_Isend}(sbuf[x], \mathtt{size}(buf[x]), t, sreqs[x])$
  \State $msg\_sent[x] \leftarrow true$ \label{alg:line:flush-sends-end}
  \EndFor
  \Else
  \For{$v \in V_i$} \Comment{traverse local subgraph} \label{alg:line:local-counting-post-begin}
  \For{$m \in N(v) \mid |v| < |m|$} \Comment{$|parent| < |child|$}
  \State $\mathit{owner} \leftarrow G_i.get\_owner(m) \mid \forall x, owner \in \mathtt{enumerate}(targets)$ 
  \If {$\mathit{owner} == i$} \Comment{local counting} \label{alg:line:local-counting-begin}
   \For{$n \in N(v) \mid n > m$}     
    \State $if \{m,n\} \in E_i \Rightarrow \tau \mathrel{+}=1$ \label{alg:line:local-counting-end}
   \EndFor
  \Else \Comment{Remote counting}
  \If {$\mathit{owner} > i$} \Comment{enable directed remote inquiry}
  \For{$n \in N(v) \mid n > m$} \label{alg:line:post-outgoing-begin}
     \If{$\mathtt{size}(buf[x]) == SIZE \; and \; !\mathit{msg\_sent}[x]$} 
     \State $\mathtt{MPI\_Isend}(sbuf[x], SIZE, owner, sreqs[x])$
     \State $msg\_sent[x] \leftarrow true$
     \Else
     \State $sbuf[x] \leftarrow sbuf[x] \cup \{m, n\}$ \Comment{remote lookup} \label{alg:line:edge-insertion}
     \State $outgoing\_msgs \mathrel{-}= 1$ \Comment{decrement counts} \label{alg:line:post-outgoing-end}
     \EndIf
  \EndFor
  \EndIf
  \EndIf
  \EndFor
  \EndFor
  \EndIf
  \State $sends\_done \leftarrow \mathtt{MPI\_Testsome}(|targets|, sreqs)$ \Comment{test completion} \label{alg:line:test-posted-isends}
  \State{$\forall c \in sends\_done \Rightarrow sbuf[c] \leftarrow \Phi, \; msg\_sent[c] \leftarrow false$} \label{alg:line:local-counting-post-end} 
  \State \Comment{{\bf \textcolor{blue}{--- Process incoming messages ---}}}
  \For{$y, s \in \mathtt{enumerate}(sources)$} \label{alg:line:process-incoming-begin}
  \If{$\mathtt{MPI\_Test}(rreqs[y])$} \Comment{receive buffer accessed if test is true}
  \State $\forall \{m,n\} \in rbuf[y] \mid If \{m,n\} \in E_i \Rightarrow \tau \mathrel{+}=1$ 
  \State $\{recv\_count[y], \; incoming\_msgs\}  \mathrel{-}= \mathtt{size}(rbuf[y])$ \label{alg:line:adjust-counts} 
  \If{$recv\_count[y] > 0$} \Comment{prepost receives incoming messages}
  \State $\mathtt{MPI\_Irecv}(rbuf[y], SIZE, s, rreqs[y])$ \label{alg:line:process-incoming-end}
  \EndIf
  \EndIf
  \EndFor
  \State \Comment{{\bf \textcolor{blue}{--- Test For completion ---}}}
  \If{$bar$} \Comment{is nonblocking barrier activated?}
  \If {$\mathtt{MPI\_Test}(nb\_bar)$} \Comment{test nonblocking barrier for completion} \label{alg:line:test-nbar}
  \State $done \leftarrow true$
  \EndIf
  \Else 
  \If{$incoming\_msgs == 0$} \Comment{processed all outstanding messages}
  \State $\mathtt{MPI\_Ibarrier}(nb\_bar)$ \Comment{activate nonblocking barrier} \label{alg:line:activate-nbar}
  \State $bar \leftarrow true$ 
  \EndIf
  \EndIf
  \EndWhile
\end{algorithmic}
}
\end{algorithm}
\subsubsection{Process incoming messages and test completion}\label{sssec:alg-incoming-completion}
Every process accepts messages from its neighboring processes (i.e., \emph{sources}); we check the status of the incoming messages (which consists of prospective edges) as shown in Lines ~\ref{alg:line:process-incoming-begin}--\ref{alg:line:process-incoming-end} and search for them in the locally owned vertex neighborhoods. Each of these prospective edges if exists closes a particular wedge, indicating the presence of a distinct triangle. We utilize MPI nonblocking barrier (see Line~\ref{alg:line:activate-nbar}) instead of a more expensive collective operation as a stopping criteria to exit from the super-step; each process invokes the nonblocking barrier after processing all the incoming messages. Testing the MPI request handle associated with the nonblocking barrier returns true when all the processes have invoked the barrier, implying completion of outstanding communication and processing. 

\subsubsection{CPU vs. GPU version}\label{sssec:alg-cpugpu}
Lines~\ref{alg:line:local-counting-begin}--\ref{alg:line:local-counting-end} corresponds to the local subgraph counting, and we develop GPU kernels for the purpose as discussed in \S\ref{ssec:method-gpu-tric} (GPU is mapped to a specific process). The GPU version copies the local subgraph from host to device memory, and allocates another $\mathcal{O}(SIZE)$ buffer for holding intermediate received data. The task of scanning the local subgraph for prospective cross edges and insertion into outgoing communication buffers (i.e., Lines~\ref{alg:line:post-outgoing-begin}--\ref{alg:line:post-outgoing-end}) is executed on the host. Lines ~\ref{alg:line:process-incoming-begin}--\ref{alg:line:process-incoming-end} are ported on GPUs (facilitating concurrent edge lookups), and prior to launching the specific kernel on a CUDA stream, the received incoming data is copied on the device. Since the data is sent from the host and processed on device, we cannot use GPU-aware MPI~\cite{wang2013gpu}. Contrastingly, the CPU version uses OpenMP \texttt{for parallel} sections for local counting (Lines ~\ref{alg:line:local-counting-begin}--\ref{alg:line:local-counting-end}) and processing incoming data (Lines ~\ref{alg:line:process-incoming-begin}--\ref{alg:line:process-incoming-end}) necessitating thread-safe MPI support.

Inspired from \textsc{TriC}~\cite{ghosh2022improved}, we compress a set of vertex pairs or edges into a sequence such that a common parent vertex is not repeated and distinct sets of neighborhoods are demarcated using a negative number (fewer communication rounds, more data packed into buffers). However, this logic also makes it difficult to effectively parallelize within the GPU blocks (since the demarcations are not known in advance). Therefore, for the GPU versions, we trade-off communication with faster local processing, tolerating redundancy. For the CPU version, we additionally experiment with lossless data compression for the communication buffers, reaching $>$90\% space efficiency (\S\ref{sec:eval-mpi}). 

\subsection{GPU methods for triangle counting}\label{ssec:method-gpu-tric}
We use a hybrid approach in engaging both CPUs and GPUs for distributed-memory computations. Developing GPU kernels following the standard template of CPU-centric graph memory-access pattern corresponding to the CSR representation significantly increases the loop overhead: requiring four nested loops for the triangle counting workload. On a GPU, this leads to major per-thread task overhead, load imbalances, thread-divergence and subpar occupancy (i.e., maximum achievable concurrency). Thus, reducing the loops is mandatory towards a practical baseline. GPUs consist of blocks of threads laid out as a rectangular grid, and the thread-blocks can be further partitioned into groups known as ``tiles''; the number of threads\slash block and the number of blocks\slash grid are part of the kernel launch parameters. 

To avoid tile-level divergence and enhance memory coalescing for single counting, we filter out\slash mask the edges (at the expense of computation and intermediate memory, but better occupancy\slash coalescing) and perform further flattening (collapsing the loop over vertices and corresponding edges) before copying them to the GPU memory and adopting a \emph{grid-stride} access pattern. Consequentially, due to relatively high \#edges\slash process\slash GPU (for most large graphs) and batching of the kernels on GPU streams, individual kernels are usually unable to perfectly fill the GPU's hardware processing units (i.e., SMs), causing it to run in distinct temporal stages referred to as \emph{waves}, which correspond to the maximum resident thread blocks executing on GPU SMs concurrently. Leftover thread blocks leave some SMs or slot capacities idle, giving rise to \emph{partial waves}, delivering less throughput overall. Next, we discuss the adaptations to incorporate independent local and receiver-side counting. 

\begin{lstlisting}[label=lst:masked-edge-kernel,caption=Masked-edge based local counting kernel]
template <int tilesz, int blocksz> __global__ void masked_edge_kernel ( const int nedges, const int* d_emsk, const int* d_ind, const int* d_elst, int* d_count ) {
  /* Initializations */
  const int bid = blockIdx.x, step = gridDim.x;
  __shared__ int range[(blocksz/tilesz)*2];
  thread_block block = this_thread_block();
  thread_group warp = tiled_partition<tilesz>(block);
  const int b_tid = block.thread_rank();
  const int w_tid = warp.thread_rank();
  int* t_range = &range[(b_tid/tilesz)*2]; 
  int i0 = (b_tid/tilesz)+(blocksz/tilesz)*bid;
  int count = 0, grid_stride = step*(blocksz/tilesz);
  /* Iterate over locally owned edges */
  for (int i = i0; i < nedges; i += grid_stride) {
    if(w_tid == 0) {
      const int id = 2*i; 
      t_range[0] = d_emsk[id]; 
      t_range[1] = d_emsk[id+1];
    }
    warp.sync();
    const int e_end = t_range[0], m = t_range[1], edge_m = d_elst[m], jstart = d_ind[edge_m], jend = d_ind[edge_m+1];
    /* Distance-1: Iterate over partial vertex edge list */
    for(int n = m + 1 + w_tid; n < e_end; n += tilesz) {
      const int edge_n = d_elst[n];      
      /* Distance-2: Iterate over edge list of neighbor */ 
      for(int j = jstart; j < jend; j++) {
        if (d_elst[j] == edge_n)
          count++;
        if (_any_sync(_activemask(), d_elst[j] >= edge_n))
          break;
      }
    }
  }
#pragma unroll
  for (int j = warp.size()/2; j > 0; j /= 2)
    count += __shfl_down_sync(FULL_MASK, count, j);
  if (w_tid == 0)
    atomicAdd((int*)d_count, count);
}
\end{lstlisting}

\subsubsection{Masked-edge based local\slash independent counting}\label{sssec:method-masked-edge}
This variant is the basis of our local\slash independent counting, where each process\slash GPU counts triangles for every vertex neighborhood it owns. A simplified kernel is shown in Listing~\ref{lst:masked-edge-kernel}; the local CSR graph structure is represented by the \texttt{d\_ind} and \texttt{d\_elst} arrays (compressed vertex neighborhood depicted by \texttt{d\_emsk}), which are common in all the kernels. We use CUDA Cooperative Groups (denoted as \texttt{cg} in Listing~\ref{lst:masked-edge-kernel}) to enable flexible collective operations within partitioned tiles of threads within a block (``tile'' corresponds to a warp). GPU shared memory and thread-local registers stores frequently reused data, to reduce the volume of global memory accesses. We also employ warp-level built-in collectives to assess the exit criteria from the innermost loop and perform block-level reductions (e.g., \texttt{\_\_any\_sync} and \texttt{\_\_shfl\_down\_sync}). 
Each thread in a warp\slash tile performs distance-2 neighborhood scans (loop in Line 25 of Listing~\ref{lst:masked-edge-kernel}). Local counting is subject to variations in arbitrary vertex neighborhoods across input graphs, affecting the parallel efficiency. For instance, when the outer loop traversing over the thread-tiles\slash warps (Line 21) is smaller than the innermost loop's trip-count, a portion of the thread-tiles\slash block are being idle, leading to underutilized SMs and reduced throughput. On the other hand, splitting distance-1 and distance-2 neighborhoods into separate kernels may improve the load imbalance, but can increase the execution times and launch overhead.


\subsubsection{Split-edge based receiver-side counting}\label{sssec:method-split-edge}
\begin{figure}[!ht]
\centering
\includegraphics[width=0.5\textwidth]{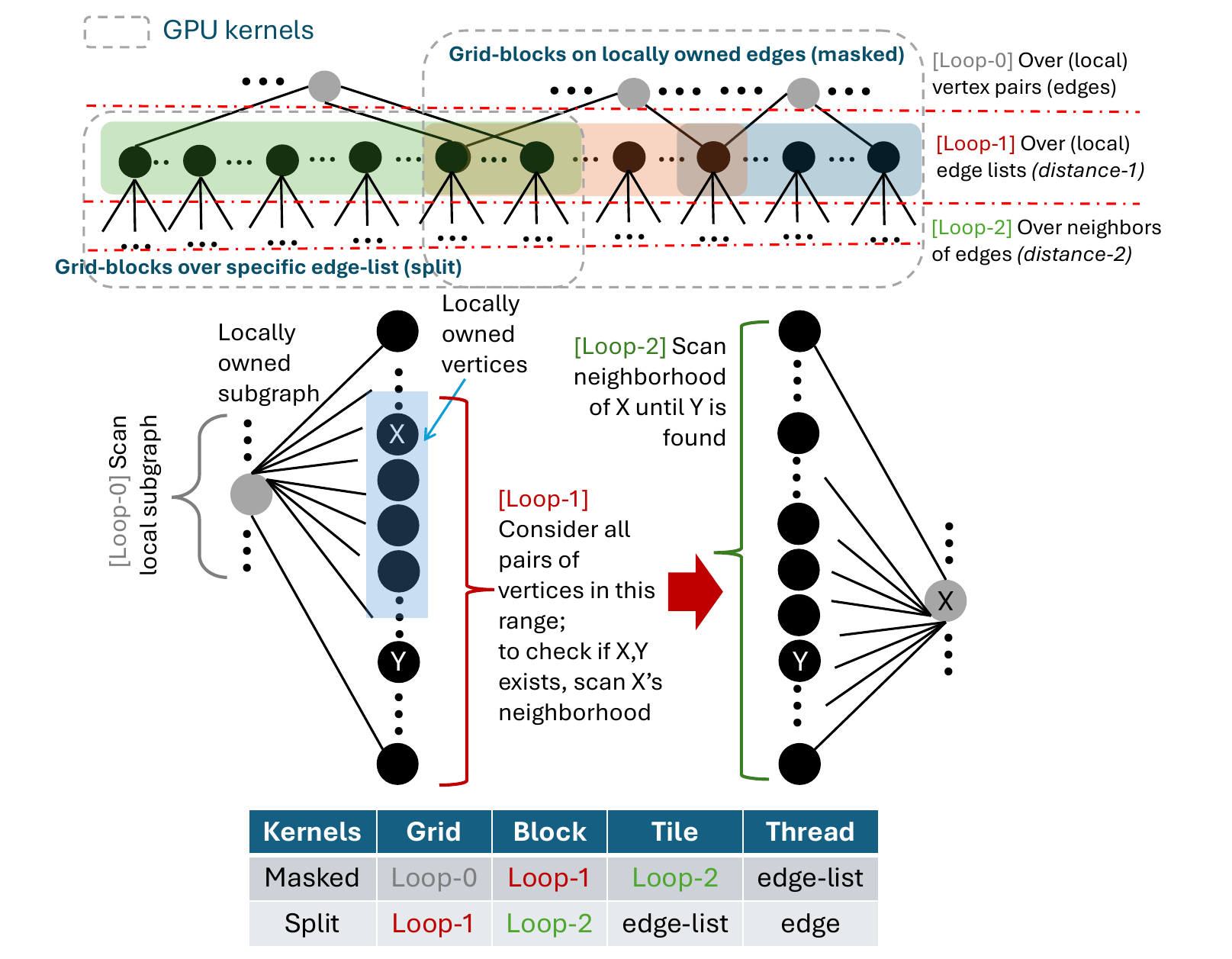}
\caption{\small Grid-block-tile level operations for masked and split-edge based triangle counting.}
\label{fig:count-triangle-split} 
\end{figure}
We use a variant of the masked-edge kernel on the receiver-side to process incoming edges (see Lines~\ref{alg:line:process-incoming-begin}--\ref{alg:line:process-incoming-end} of Algorithm~\ref{alg:tric-algorithm}), invoked concurrently over CUDA streams. This is equivalent to peeling the outermost loop (over masked-edges), and assigning the iterations of the partial edge-list (distance-1) over GPU grid-blocks and the nested loop of the distance-2 neighborhood is processed by the thread-tiles\slash block (each thread executes an iteration of the innermost loop, instead of the entire loop). Consequently, split-edge kernel roughly corresponds to Line 22 onward in Listing~\ref{lst:masked-edge-kernel}. Fig.~\ref{fig:count-triangle-split} depicts the scheduling the loop-nests across grid, block and thread-tiles of a GPU for masked and split kernels. Reuse potential of vertex ranges are not known in advance; prefetching arbitrary (distance-2) ranges into faster shared-memory cannot guarantee performance (thereby amortizing global-shared-memory copies) if the range is accessed sparingly.

\subsubsection{GPU Inter-Process Communication}\label{sssec:method-ipc}
GPU Inter-Process Communication (IPC) enables different processes within the same node to share GPU memory pointers, allowing peer-to-peer access to the memory mapped to a remote GPU (assuming GPUs to be on the same PCIe switch or interconnected via proprietary NVLink~\regtm{}), bypassing explicit data movement within a node (modern architectures expand this concept to NVLink domains, beyond node boundaries~\cite{tirumala2024nvidia}). We leverage IPC in our distributed-memory and single node implementations, performing a one-time \texttt{all-gather} operation to share the IPC handles within a node. 

IPC variants further segregates edges (within a compute node) based on whether they belong to local or remote vertex neighborhoods (from the point-of-view of a process\slash device, see Fig.~\ref{fig:counting-precedence}), requiring intermediate buffers to maintain non-consecutive list of edges. Thus, single-node IPC variants invoke masked-edge kernels (\S\ref{sssec:method-masked-edge}) on two distinct sets of edges (i.e., local and remote: IPC is only applicable on the remote set, which is further separated by target processes), whereas the MPI+IPC variant additionally invokes the split-edge kernels (\S\ref{sssec:method-split-edge}) for processing incoming edges. Each process invokes several kernels\slash GPU concurrently on streams. Respective volumes of the local and remote vertex sets determine the thread-block-grid on which a particular kernel is executed. Inherent differences in the structure of the graphs, resultant partitions and working-sets can thus lead to disparities in the number of concurrent thread blocks used in executing the kernels.
\section{Evaluation}\label{sec:eval}
In this section, we analyze our distributed-memory triangle counting implementations on CPUs and GPUs. 
\subsection{Datasets\slash Platforms}\label{sec:eval-setup}
\subsubsection{Datasets}\label{ssec:eval-setup-datasets}
We evaluate our implementations on a diverse suite of real-world graph datasets spanning citation networks, social networks, and large-scale web graphs listed in Table~\ref{tab:datasets} from SuiteSparse~\cite{kolodziej2019suitesparse} and GraphChallenge~\cite{samsi2018graphchallenge} repositories. 
\begin{table}[!ht]
\centering
\caption{\small Graphs used in single and multi-node GPU evaluations.}\label{tab:datasets}
{\scriptsize
\begin{tabular}{|lllll|}
\hline
\multicolumn{1}{|l|}{\textbf{Graphs}}    & \multicolumn{1}{l|}{\textbf{\#Vertices}}  & \multicolumn{1}{l|}{\textbf{\#Edges}}   & \multicolumn{1}{l|}{\textbf{Max Edges}} & \textbf{\#Triangles}    \\ \hline \hline
\multicolumn{5}{|c|}{\textbf{MPI/IPC evaluations (multi-GPU\slash multi-node)}}                                                                                                                \\ \hline \hline
\multicolumn{1}{|l|}{orkut}              & \multicolumn{1}{l|}{3M}     & \multicolumn{1}{l|}{234.3M}    & \multicolumn{1}{l|}{29K}       & 627,584,181     \\ \hline
\multicolumn{1}{|l|}{arabic-2005}        & \multicolumn{1}{l|}{22.7M}    & \multicolumn{1}{l|}{639.9M}    & \multicolumn{1}{l|}{2.5M}        & 24,590,348,965  \\ \hline
\multicolumn{1}{|l|}{webbase-2001}       & \multicolumn{1}{l|}{118.1M}   & \multicolumn{1}{l|}{1.9B}  & \multicolumn{1}{l|}{254.9M}      & 24,510,198,252  \\ \hline
\multicolumn{1}{|l|}{friendster}         & \multicolumn{1}{l|}{65.6M}    & \multicolumn{1}{l|}{3.6B}  & \multicolumn{1}{l|}{451.5M}      & 4,173,724,142   \\ \hline
\multicolumn{1}{|l|}{uk2007-05}          & \multicolumn{1}{l|}{105.8M}   & \multicolumn{1}{l|}{6.6B}  & \multicolumn{1}{l|}{825.5M}      & 286,701,284,103 \\ \hline
\multicolumn{1}{|l|}{clueweb09}          & \multicolumn{1}{l|}{4.7B} & \multicolumn{1}{l|}{7.9B}  & \multicolumn{1}{l|}{992.4M}      & 83,434,096      \\ \hline
\multicolumn{1}{|l|}{twitter7}           & \multicolumn{1}{l|}{41.6M}    & \multicolumn{1}{l|}{2.9B}  & \multicolumn{1}{l|}{367.5M}      & 105,717,152,493 \\ \hline
\multicolumn{1}{|l|}{sk-2005}            & \multicolumn{1}{l|}{50.6M}    & \multicolumn{1}{l|}{3.8B}  & \multicolumn{1}{l|}{487.3M}      & 192,464,185,701 \\ \hline
\multicolumn{1}{|l|}{it-2004}            & \multicolumn{1}{l|}{41.2M}    & \multicolumn{1}{l|}{2.2B}  & \multicolumn{1}{l|}{287.6M}      & 97,408,315,364  \\ \hline
\multicolumn{1}{|l|}{GAP-kron}           & \multicolumn{1}{l|}{134.2M}   & \multicolumn{1}{l|}{4.2B}  & \multicolumn{1}{l|}{527.9M}      & 106,873,365,648 \\ \hline
\multicolumn{1}{|l|}{AGATHA-2015}        & \multicolumn{1}{l|}{183.9M}   & \multicolumn{1}{l|}{11.5B} & \multicolumn{1}{l|}{1.4B}    & 1,314,062,287   \\ \hline
\multicolumn{1}{|l|}{scale18}            & \multicolumn{1}{l|}{174.1K}       & \multicolumn{1}{l|}{7.6M}      & \multicolumn{1}{l|}{951.9K}          & 82,287,285      \\ \hline
\multicolumn{1}{|l|}{scale19}            & \multicolumn{1}{l|}{335.3K}       & \multicolumn{1}{l|}{15.4M}     & \multicolumn{1}{l|}{1.9M}        & 186,288,972     \\ \hline
\multicolumn{1}{|l|}{scale20}            & \multicolumn{1}{l|}{645.8K}       & \multicolumn{1}{l|}{31.3M}     & \multicolumn{1}{l|}{3.9M}        & 419,349,784     \\ \hline
\multicolumn{1}{|l|}{scale21}            & \multicolumn{1}{l|}{1.2M}     & \multicolumn{1}{l|}{63.4M}     & \multicolumn{1}{l|}{7.9M}        & 935,100,883     \\ \hline
\multicolumn{1}{|l|}{scale22}            & \multicolumn{1}{l|}{2.3M}     & \multicolumn{1}{l|}{128.1M}    & \multicolumn{1}{l|}{16M}       & 2,067,392,370   \\ \hline
\multicolumn{1}{|l|}{scale23}            & \multicolumn{1}{l|}{4.6M}     & \multicolumn{1}{l|}{258.5M}    & \multicolumn{1}{l|}{32.3M}       & 4,549,133,002   \\ \hline
\multicolumn{1}{|l|}{scale24}            & \multicolumn{1}{l|}{8.8M}     & \multicolumn{1}{l|}{520.5M}    & \multicolumn{1}{l|}{65M}       & 9,936,161,560   \\ \hline
\multicolumn{1}{|l|}{scale25}            & \multicolumn{1}{l|}{17M}    & \multicolumn{1}{l|}{1B}  & \multicolumn{1}{l|}{130.8M}      & 21,575,375,802  \\ \hline \hline
\multicolumn{5}{|c|}{\textbf{Single GPU evaluations}}                                                                                                                         \\ \hline \hline
\multicolumn{1}{|l|}{as20000102}         & \multicolumn{1}{l|}{6.4K}         & \multicolumn{1}{l|}{25.1K}         & \multicolumn{1}{l|}{1.4K}            & 6,584           \\ \hline
\multicolumn{1}{|l|}{ca-GrQc}            & \multicolumn{1}{l|}{5.2K}         & \multicolumn{1}{l|}{28.9K}         & \multicolumn{1}{l|}{81}               & 48,260          \\ \hline
\multicolumn{1}{|l|}{oregon2\_010512}    & \multicolumn{1}{l|}{11.2K}        & \multicolumn{1}{l|}{62.6K}         & \multicolumn{1}{l|}{2.4K}            & 72,866          \\ \hline
\multicolumn{1}{|l|}{as-caida20071105}   & \multicolumn{1}{l|}{26.4K}        & \multicolumn{1}{l|}{106.7K}        & \multicolumn{1}{l|}{2.6K}            & 36,365          \\ \hline
\multicolumn{1}{|l|}{facebook} & \multicolumn{1}{l|}{4K}         & \multicolumn{1}{l|}{176.4K}        & \multicolumn{1}{l|}{1K}            & 1,612,010       \\ \hline
\multicolumn{1}{|l|}{ca-CondMat}         & \multicolumn{1}{l|}{23.1K}        & \multicolumn{1}{l|}{186.8K}        & \multicolumn{1}{l|}{279}              & 173,361         \\ \hline
\multicolumn{1}{|l|}{p2p-Gnutella31}     & \multicolumn{1}{l|}{62.5K}        & \multicolumn{1}{l|}{295.7K}        & \multicolumn{1}{l|}{95}               & 2,024           \\ \hline
\multicolumn{1}{|l|}{email-Enron}        & \multicolumn{1}{l|}{36.6K}        & \multicolumn{1}{l|}{367.6K}        & \multicolumn{1}{l|}{1.3K}            & 727,044         \\ \hline
\multicolumn{1}{|l|}{ca-AstroPh}         & \multicolumn{1}{l|}{18.7K}        & \multicolumn{1}{l|}{396.1K}        & \multicolumn{1}{l|}{504}              & 1,351,441       \\ \hline
\multicolumn{1}{|l|}{loc-brightkite}     & \multicolumn{1}{l|}{58.2K}        & \multicolumn{1}{l|}{428.1K}        & \multicolumn{1}{l|}{1.1K}            & 494,728         \\ \hline
\multicolumn{1}{|l|}{cit-HepTh}          & \multicolumn{1}{l|}{27.7K}        & \multicolumn{1}{l|}{704.5K}        & \multicolumn{1}{l|}{2.4K}            & 1,478,735       \\ \hline
\multicolumn{1}{|l|}{email-EuAll}        & \multicolumn{1}{l|}{265.2K}       & \multicolumn{1}{l|}{728.9K}        & \multicolumn{1}{l|}{7.6K}            & 267,313         \\ \hline
\multicolumn{1}{|l|}{soc-Epinions1}      & \multicolumn{1}{l|}{75.8K}        & \multicolumn{1}{l|}{811.4K}        & \multicolumn{1}{l|}{3K}            & 1,624,481       \\ \hline
\multicolumn{1}{|l|}{cit-HepPh}          & \multicolumn{1}{l|}{34.5K}        & \multicolumn{1}{l|}{841.7K}        & \multicolumn{1}{l|}{846}              & 1,276,868       \\ \hline
\multicolumn{1}{|l|}{soc-Slashdot0902}   & \multicolumn{1}{l|}{82.1K}        & \multicolumn{1}{l|}{1M}      & \multicolumn{1}{l|}{2.5K}            & 602,592         \\ \hline
\multicolumn{1}{|l|}{amazon0302}         & \multicolumn{1}{l|}{262.1K}       & \multicolumn{1}{l|}{1.7M}      & \multicolumn{1}{l|}{420}              & 717,719         \\ \hline
\multicolumn{1}{|l|}{loc-gowalla}        & \multicolumn{1}{l|}{196.5K}       & \multicolumn{1}{l|}{1.9M}      & \multicolumn{1}{l|}{14.7K}           & 2,273,138       \\ \hline
\multicolumn{1}{|l|}{U1a}                & \multicolumn{1}{l|}{67.7M}    & \multicolumn{1}{l|}{138.7M}    & \multicolumn{1}{l|}{35}               & 325             \\ \hline
\multicolumn{1}{|l|}{V2a}                & \multicolumn{1}{l|}{55M}    & \multicolumn{1}{l|}{117.2M}    & \multicolumn{1}{l|}{39}               & 1,443           \\ \hline
\end{tabular}
}
\end{table}
\subsubsection{Platforms}\label{ssec:eval-setup-platforms}
We use three distinct platforms: 1) \emph{CPU\slash GPU evaluations}: NERSC Perlmutter supercomputer consisting of 3,072 CPU-only compute nodes and 1,792 GPU-accelerated nodes. Each node uses two 64-core AMD EPYC 7763 CPUs (128 cores / 256 threads) with 512GB of DDR4 memory, organized into eight NUMA domains, and is connected to the Slingshot 11 interconnect via a single Network Interface (NIC) card. The GPU nodes have a single 64-core AMD EPYC 7763 CPU and 4$\times$ NVIDIA A100 GPUs\slash node (third-generation NVLink connecting GPUs\slash node: 600GB/s GPU-GPU b/w) with 40\slash 80GB of HBM2(e) memory (we use 40GB A100), 256GB of DDR4 memory\slash node and 4$\times$ HPE Slingshot 11 NICs. We use PrgEnv-nvidia/8.6.0 (with cudatoolkit/12.9) and cray-mpich/9.0.1, and CrayPAT\slash Perftools (perftools-base/25.0.9) for profiling. 2) \emph{Distributed-memory GPU evaluations}: We use ALCF Polaris with 560 nodes comprising of two 64-core (128 cores / 256 threads) AMD EPYC 7713 CPU and 4$\times$ NVIDIA A100 GPUs\slash node (third-generation NVLink connections) with 40GB of HBM2 memory, 512GB of DDR4 memory\slash node (2$\times$ of Perlmutter, consequently we use 2$\times$ larger communication buffers) and 4$\times$ HPE Slingshot 11 NICs. We use PrgEnv-nvidia/8.6.0 (with cudatoolkit/12.9) and cray-mpich/9.0.1 on Polaris platform. 3) \emph{Single-node NVIDIA DGX A100\slash H100}: We conduct single GPU\slash single-node multi-GPU experiments on NVIDIA DGX\regtm{} GPU server platforms (consisting of 8$\times$ A100\slash H100 GPUs with 40\slash 80GB HBM2(e)\slash 3 memory) and two AMD EPYC 7742 CPUs (128 cores / 256 threads). Our DGX A100 platform includes third-generation NVLink GPU interconnect, whereas DGX H100 includes fourth-generation NVLink (up to 900GB/s b/w). We use CUDA/12.3, GCC/12.2.0 and OpenMPI/5.0.5 for building codes on the DGX platforms. NVIDIA H100 is a successor of the A100 GPU, and have 20\% more SMs (about twice the \#CUDA cores) with 30\% higher clock speed\slash relative to A100, indicating potential for ``free performance''~\cite{choquette2023nvidia}.  

\subsubsection{Results interpretation}\label{ssec:eval-setup}
We capture the load imbalance owing to repetitive and variable distance-2 scans by overlaying the average local computation with min--max range on the total computation\slash communication times. The total time reported includes local counting, preparing outgoing\slash \texttt{send} communication buffers (for scanning neighborhood of remotely owned vertices), initiating communication, processing incoming communication and performing local neighborhood scans for the vertices in the incoming (or \texttt{recv}) buffer (essentially Algorithm~\ref{alg:tric-algorithm}, except graph loading\slash CSR construction times which is about 1--5\% of the total time). 

\subsubsection{Variants}\label{ssec:eval-variants} Aside from baseline distributed-memory evaluations on CPUs and GPUs, we have developed specialized variants to optimize and avoid communication. To optimize communication, we consider lossless data compression (for CPU variants) and Inter-Process Communication (IPC, using CUDA IPC to enable ``zero copy'' leveraging proprietary GPU interconnects such as NVLink\regtm{}) for the GPU variants (pure IPC for single-node multi-GPU while mixed-IPC\slash within node and communication\slash across nodes for distributed-memory GPU). Since using available compute resources can drive node bandwidth, we also develop a hybrid variant using CPUs (with multithreading) and GPUs simultaneously. The motivations behind these variants are driven by trade-offs: relative to CPU-only, GPU-only distributed-memory implementations can exhibit non-trivial load imbalances despite better overall performances; hybrid CPU\slash GPU implementations on the other hand can reduce local\slash independent counting costs, but increases the communication overheads. 

\subsection{Scale-Out CPU evaluations}\label{sec:eval-mpi}
\begin{figure}[!ht]
\centering
\includegraphics[width=0.9\linewidth]{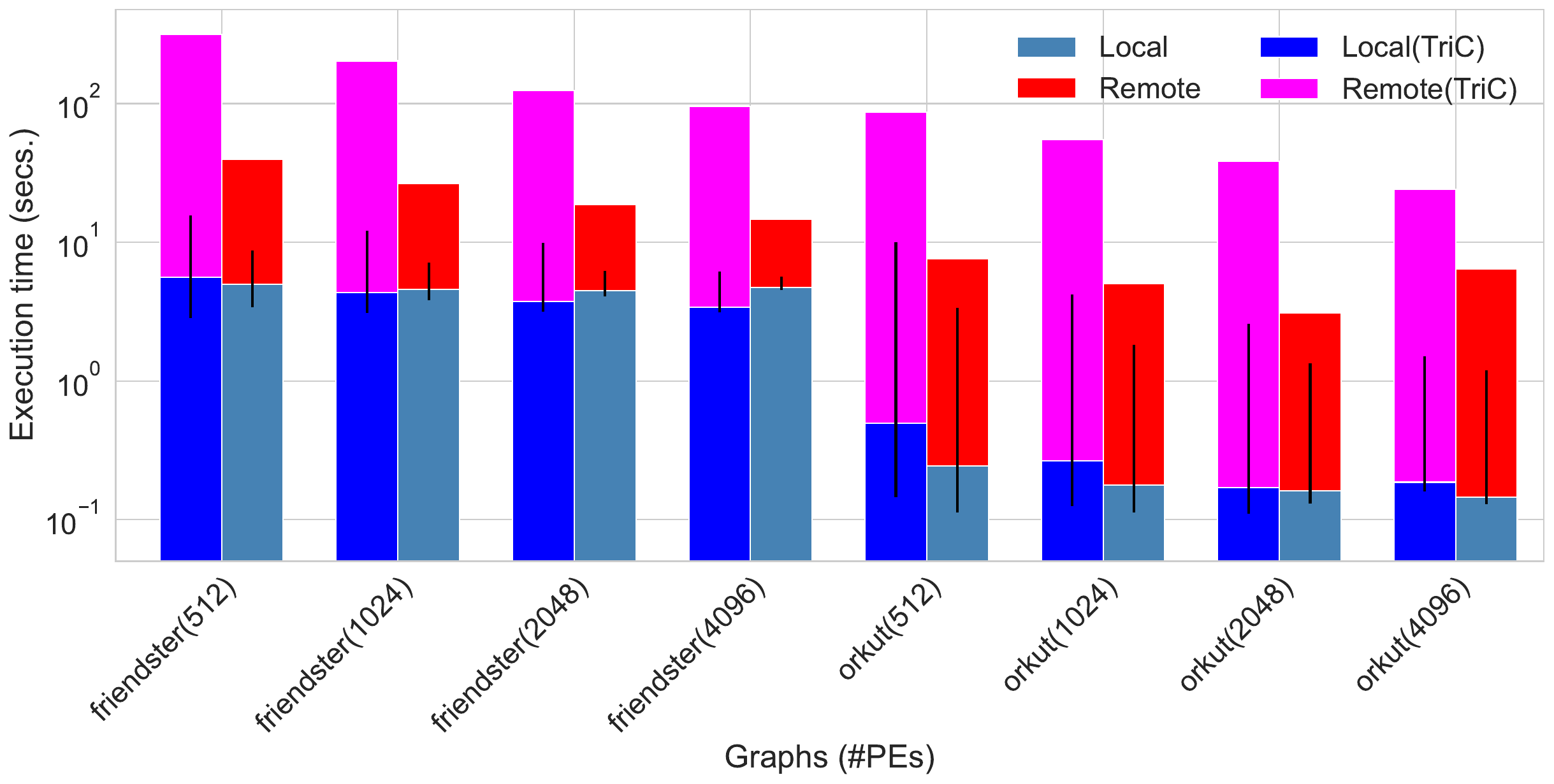}
\caption{\small Comparing our distributed-memory (CPU) implementation with TriC on NERSC Perlmutter for friendster and orkut graphs.}
\label{fig:xxtric-vs-tric-cpu} 
\end{figure}

 We compare our CPU baseline with \textsc{TriC}~\cite{ghosh2022improved}, an MPI-based CPU-only triangle counting implementation that adopts a similar enumeration-heavy approach as ours (using compacted buffered non-blocking sends\slash receives), but entails triple-counting (limited filtering of wedges), increasing load imbalance and communication volumes as shown in Fig.~\ref{fig:xxtric-vs-tric-cpu}. 
 We use the same buffer sizes for evaluating our implementation against TriC for orkut and friendster inputs on 16--128 nodes (Processes-Per-Node, PPN=32) of NERSC Perlmutter, and observe an order of magnitude improvement in the end-to-end times and at least 50\% less load imbalance, owing to enhanced filtering. 
Compression is a straightforward optimization to reduce communication volumes at the expense of CPU or out-of-core processing, and recent works have shown nontrivial improvements for pipelined MPI collectives~\cite{ramesh2022designing}. 
\begin{figure}[!ht]
\centering
\includegraphics[width=0.9\linewidth]{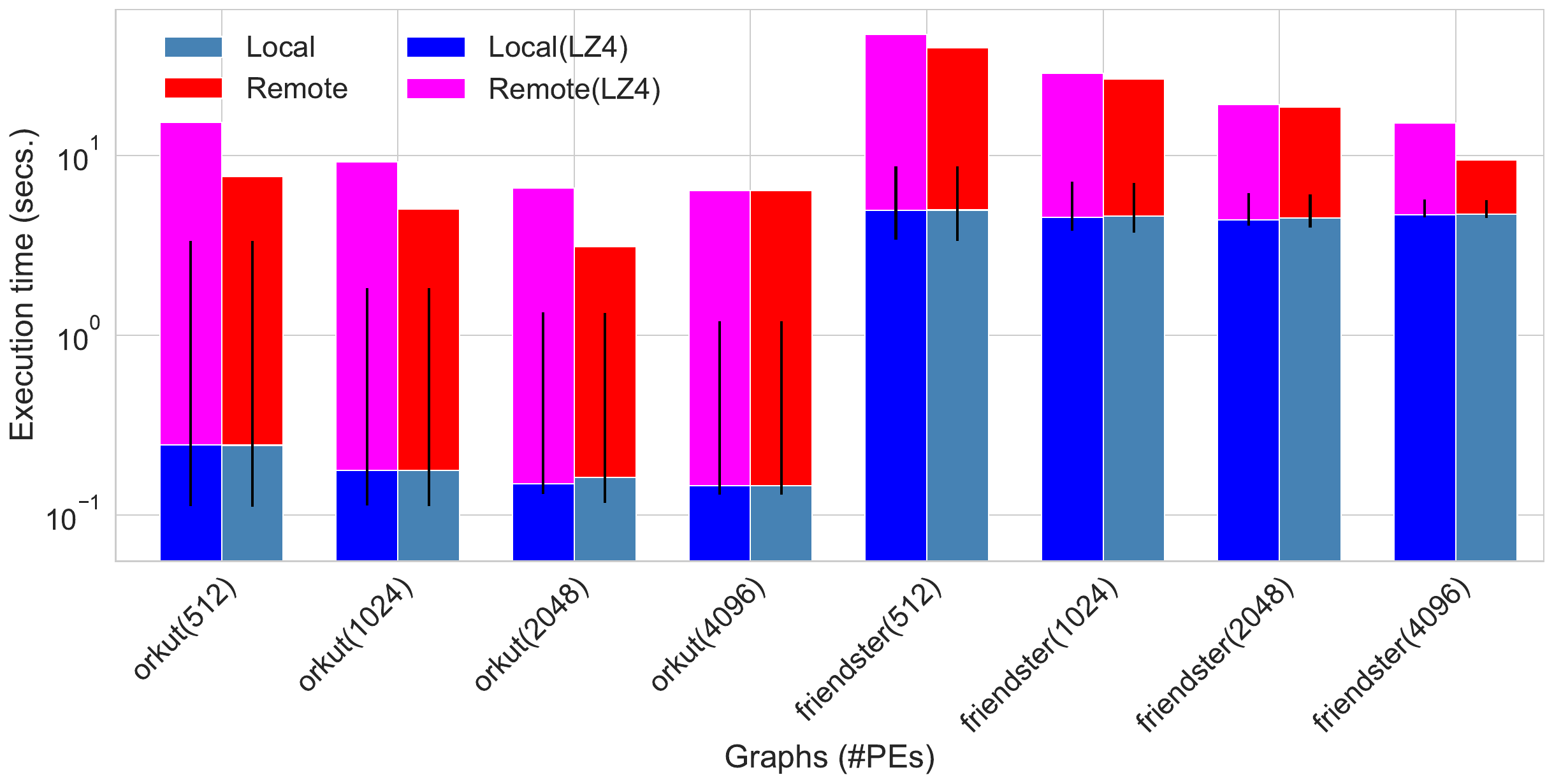}
\caption{\small Impact of LZ4 compression (sender) and decompression (receiver) on our CPU baseline for friendster and orkut graphs.}
\label{fig:xxtric-lz4-cpu} 
\end{figure}
To reduce the overall communication volume, we experimented with compressing the communication buffers using the fast LZ4 algorithm (achieves lossless compression by encoding repeated byte-sequences in the stream)~\cite{ziv2003universal}. Due to memory overheads of compression, we use smaller buffer sizes relative to the CPU-baseline evaluations in Fig.~\ref{fig:xxtric-vs-tric-cpu}. Despite significant compression factors (8:1--55:1: $>$90\% space saved), irregular communication offers limited possibility to amortize the relative costs, resulting in 20--50\% higher execution times as shown in Fig.~\ref{fig:xxtric-lz4-cpu}. 
\subsection{Scale-Out\slash Up GPU evaluations}\label{sec:eval-mpi-gpu}
\begin{figure}[!ht]
\centering
\includegraphics[width=0.9\linewidth]{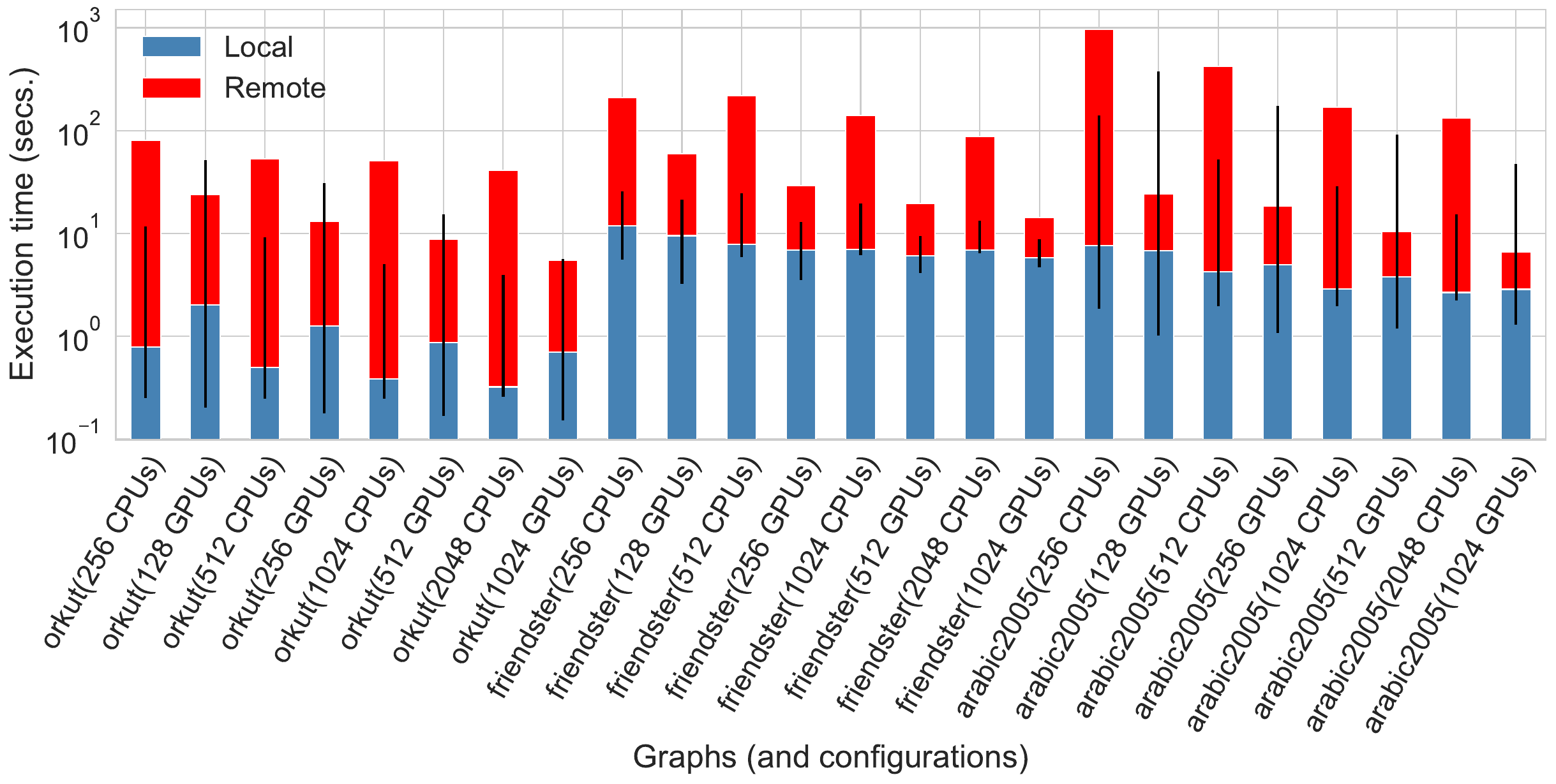}
\caption{\small CPU (odd bars) vs. GPU (even bars) on NERSC Perlmutter.}
\label{fig:perlmutter-cpu-gpu} 
\end{figure}

\paragraph*{CPU vs. GPU} It is customary to contrast distributed-memory performance between CPUs and GPUs, and Fig.~\ref{fig:perlmutter-cpu-gpu} makes this comparison on NERSC Perlmutter platform (with distinct CPU and GPU partitions).
\begin{figure*}[ht!]
    \centering
    \begin{subfigure}[t]{0.4\textwidth}
        \centering
        \includegraphics[width=\textwidth]{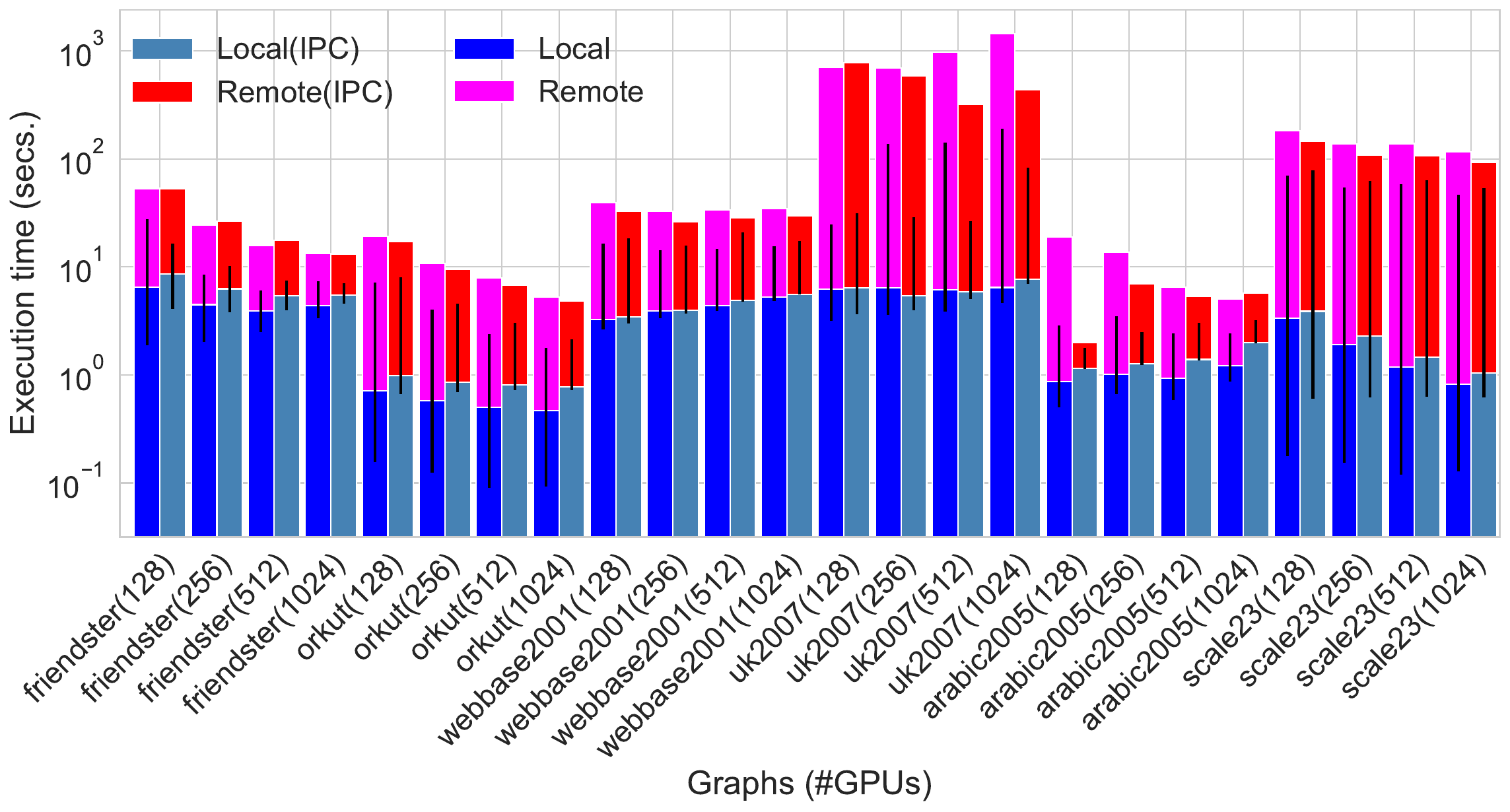}
        \caption{\small NERSC Perlmutter.}
        \label{fig:perlmutter-gpu-ipc-perf}
    \end{subfigure}%
    ~ 
     \begin{subfigure}[t]{0.53\textwidth}
         \centering
         \includegraphics[width=\textwidth]{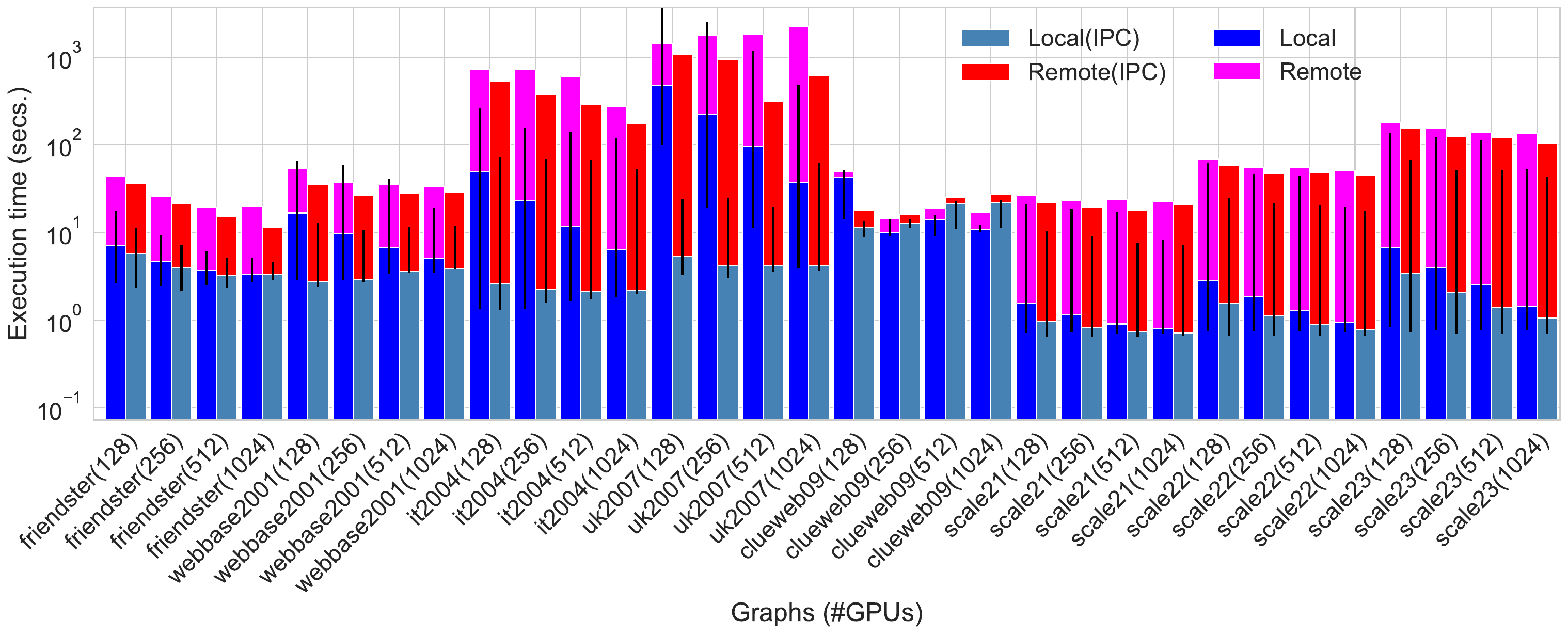}
        \caption{\small ALCF Polaris.}
        \label{fig:polaris-gpu-ipc-perf}
    \end{subfigure}%
    \caption{\small Execution times (secs.) of MPI (odd bars) and MPI+IPC (even bars) GPU variants with local\slash remote breakup.}
    \label{fig:polaris-gpu-ipc-perf} 
\end{figure*}

In Fig.~\ref{fig:perlmutter-cpu-gpu}, consecutive bars represent CPU and GPU variants, with different communication buffer sizes and PPN (=\#GPUs(4) for GPU and =32 for CPU), keeping the \# compute nodes consistent. CPU variants further uses 4 OpenMP threads\slash process to accelerate distance 1\slash 2 scans. In terms of relative execution times, GPU variants are about 3--20$\times$ faster for diverse inputs. 

\begin{figure}[ht!]
    \centering
    \begin{subfigure}[t]{0.5\linewidth}
        \centering
        \includegraphics[width=\linewidth]{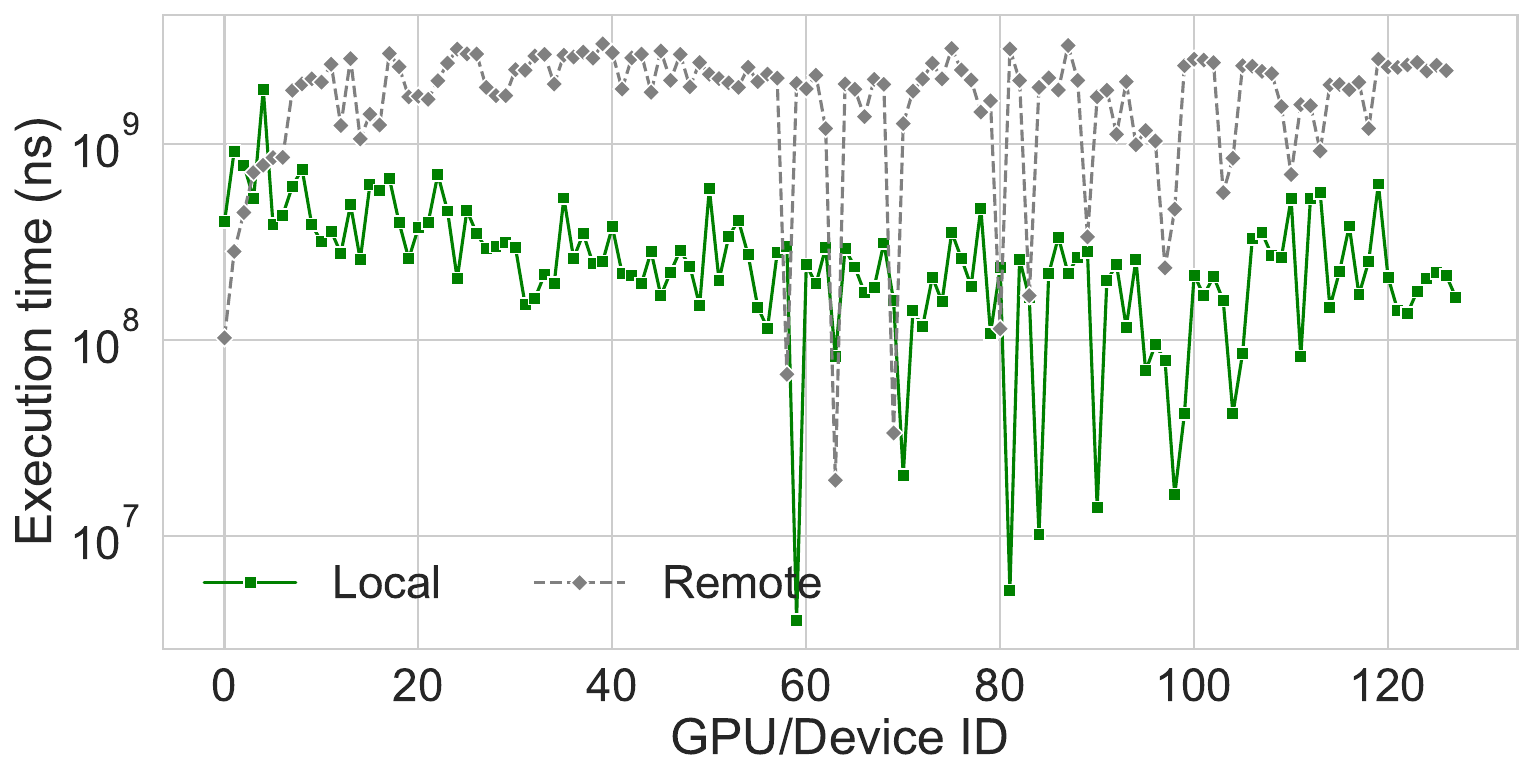}
        \caption{\small Local\slash Remote for friendster.}
        \label{fig:nsys-masked-fster}
    \end{subfigure}%
    ~ 
     \begin{subfigure}[t]{0.5\linewidth}
         \centering
         \includegraphics[width=\linewidth]{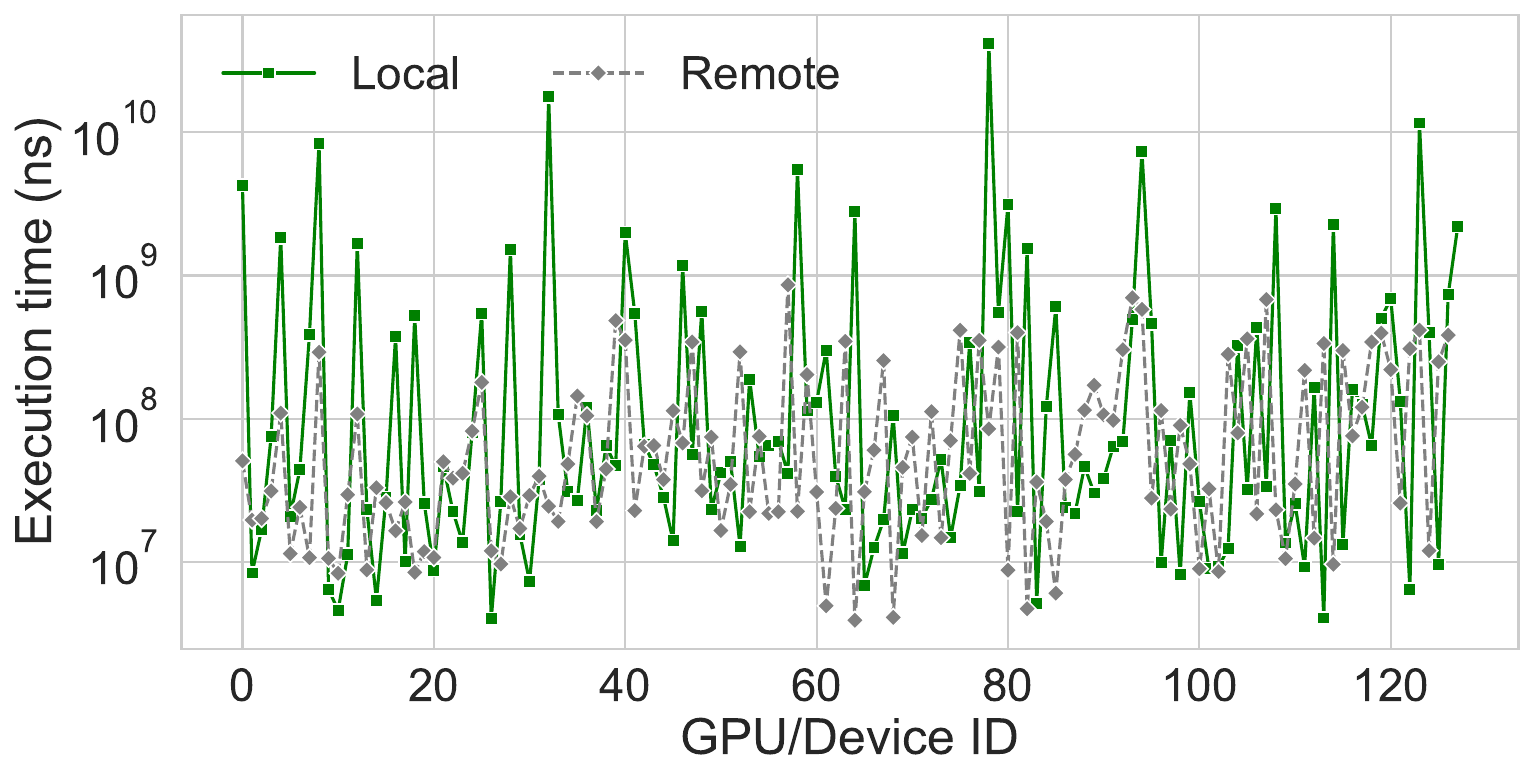}
        \caption{\small Local\slash Remote for orkut.}
        \label{fig:nsys-masked-orkut}
    \end{subfigure}%
    \caption{Times (ns)\slash GPU for 128 GPUs on Perlmutter.}
    \label{fig:kernel-times-gpus-perlmutter} 
\end{figure}
However, local\slash independent counting between the CPU and GPU variants are on par (due to additional intermediate edge-masking for the GPU variants, see \S\ref{sssec:method-masked-edge}), but, GPU variants largely surpass the CPU baselines at receiver-side processing over concurrent CUDA streams. Since the GPU variants use low PPNs, they can accommodate larger per-process communication buffers---trading-off communication frequency with faster GPU processing of split kernels (see \S\ref{sssec:method-split-edge}), and, overall less kernel invocations. Load imbalance is non-trivial across the input configurations (despite being subdued relative to the CPU variants and associated scalability), as captured on Fig.~\ref{fig:kernel-times-gpus-perlmutter} for friendster and orkut.
\begin{figure}[!ht]
\centering
\includegraphics[width=\linewidth]{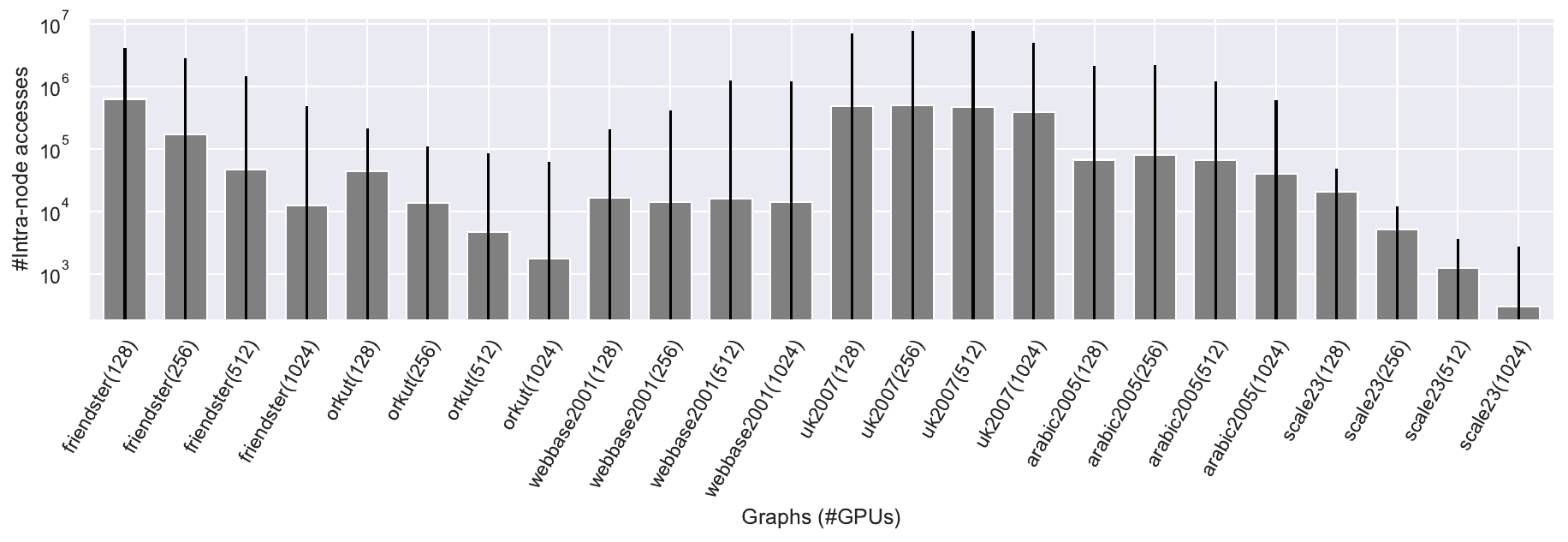}
\caption{\small Average \#node-local enumerations on NERSC Perlmutter.}
\label{fig:local-accesses-perlmutter} 
\end{figure}
\begin{figure}[!ht]
\centering
\includegraphics[width=\linewidth]{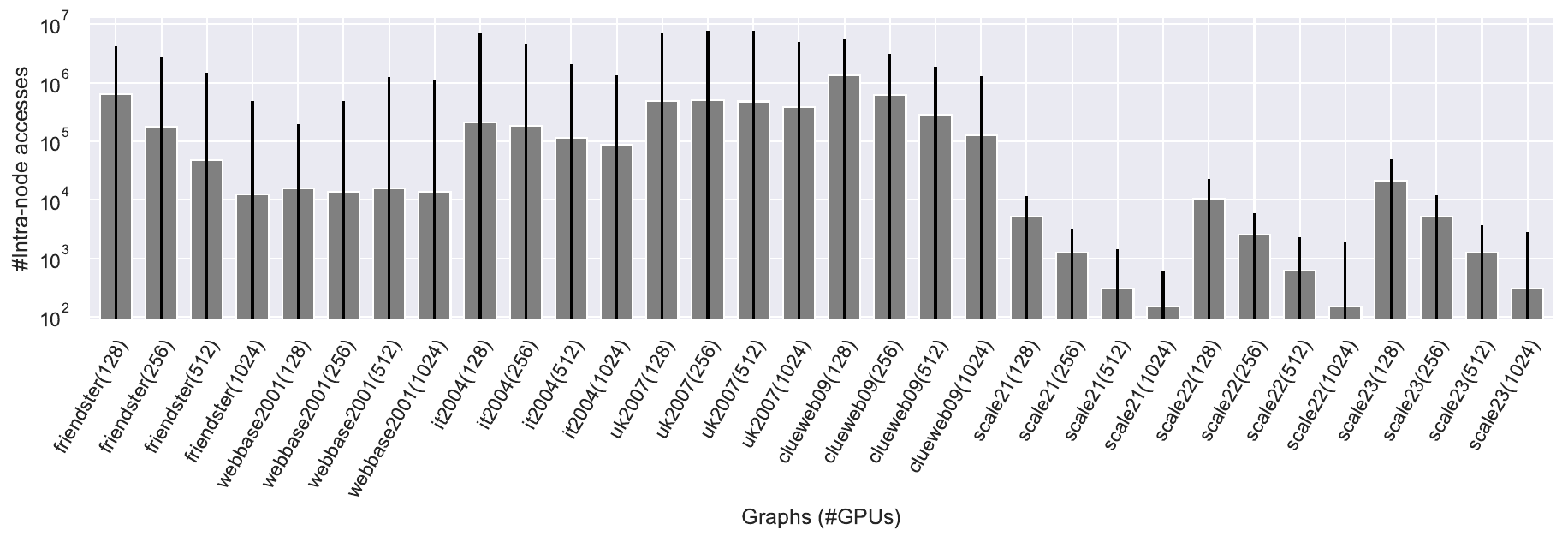}
\caption{\small Average \#node-local enumerations on ALCF Polaris.}
\label{fig:local-accesses-polaris} 
\end{figure}
\paragraph*{Mitigating load imbalance} Reducing variable communication is key towards mitigating load imbalance in general. We observe nontrivial \emph{node-local enumerations} (enumeration task on one GPU trying to access a remotely owned vertex resident on another GPU within the same compute node), typically corresponding to a maximum about 0.1--10\% of the total vertices (scale-free graphs can be exceptions, explained momentarily), on our baseline distributed-memory GPU implementation, which would otherwise have to go through the MPI communication stack. Cray MPICH supports IPC for MPI intra-node transfers involving GPU-attached memory regions by default~\cite{kandalla2023designing}, but our transfers involve host buffers due to serialization in the sender-side buffering. Figures~\ref{fig:local-accesses-perlmutter} and ~\ref{fig:local-accesses-polaris} shows the volume of average node-local accesses (with min--max bars, \emph{min} is always zero) for various graphs and configurations on Perlmutter and Polaris platforms (due to 2$\times$ memory\slash node, we use 2$\times$ larger communication buffers on Polaris). Higher volumes indicate possible IPC advantage, conversely subdued for relatively lower volumes. Typically, for scale-free graphs with power-law distributions (i.e., relatively small number of ``hub'' nodes with a vast majority of connections) such as \emph{scale21--23}, \# node-local enumerations linearly diminish with increasing processes (this effect is more subdued for social network and web graphs). Consequently, combining MPI for inter-node and IPC for intra-node led to about 10--80\% better performance on Perlmutter\slash Polaris platforms as shown in Figures~\ref{fig:perlmutter-gpu-ipc-perf} and ~\ref{fig:polaris-gpu-ipc-perf}, relative to MPI-only GPU baseline. However, few exceptions indicate a reversal in load imbalance due to increasing node-local enumerations (minor impact on communication, but about 20--40\% increase in local computations with IPC): on Perlmutter 5\slash 24 cases depicted about 1--10\% slowdown, whereas on Polaris 3\slash 32 (i.e., clueweb09) show 10--60\% drop in performance (worst case: local enumeration 2$\times$ more expensive --- impact of graph structure and larger buffers). 

\begin{figure}[!ht]
\centering
\includegraphics[width=0.9\linewidth]{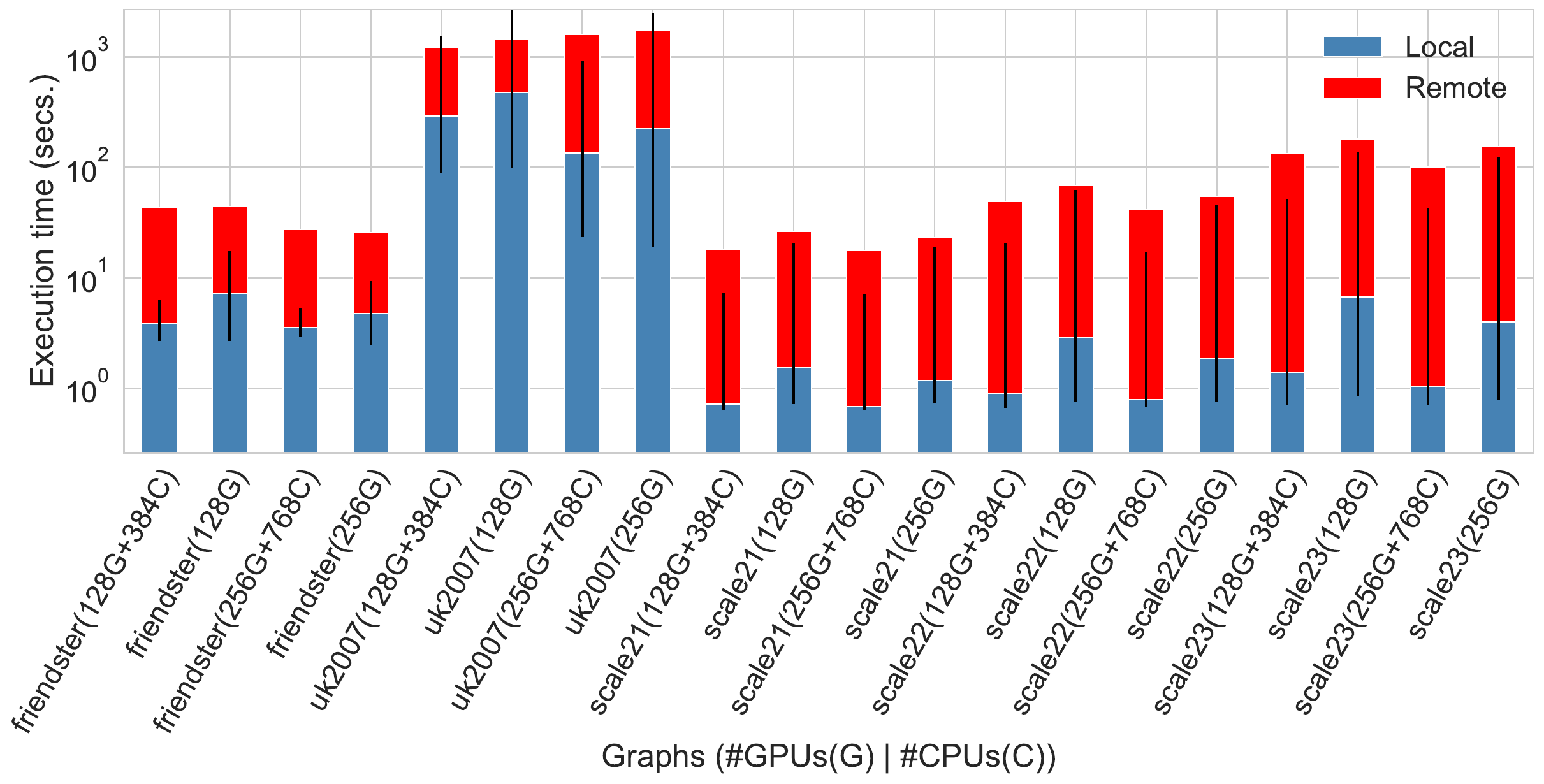}
\caption{\small Hybrid CPU and GPU variants on ALCF Polaris. \emph{Odd} bars: CPU + GPU and \emph{Even} bars: GPU only.}
\label{fig:polaris-multi-masked-perf} 
\end{figure}

Another trade-off to lessen the load imbalance considers hybrid CPU\slash GPU resources, at the expense of increasing remote communication. We explore such a hybrid CPU\slash GPU variant on Polaris, which demonstrates 3--35\% better performance relative to the baseline MPI-GPU variant, as shown in Fig.~\ref{fig:polaris-multi-masked-perf}. While the baseline and IPC variants use as many PPNs as \#GPUs (i.e., 4, see Fig.~\ref{fig:perlmutter-gpu-ipc-perf}), the hybrid variant uses 3$\times$ \#processes (PPN=16: 4(GPUs) + 12(CPUs)) and 2$\times$ smaller communication buffers (thereby increasing both communication frequency and volume). Specifically, scale-free graphs depict 24--35\% improvements relative to the baseline MPI, and 10--20\% better performance than reported MPI+IPC in Fig.~\ref{fig:polaris-gpu-ipc-perf}.   

\subsection{Single-node evaluations}\label{sec:eval-ipc}
\begin{figure}[!ht]
\centering
\includegraphics[width=\linewidth]{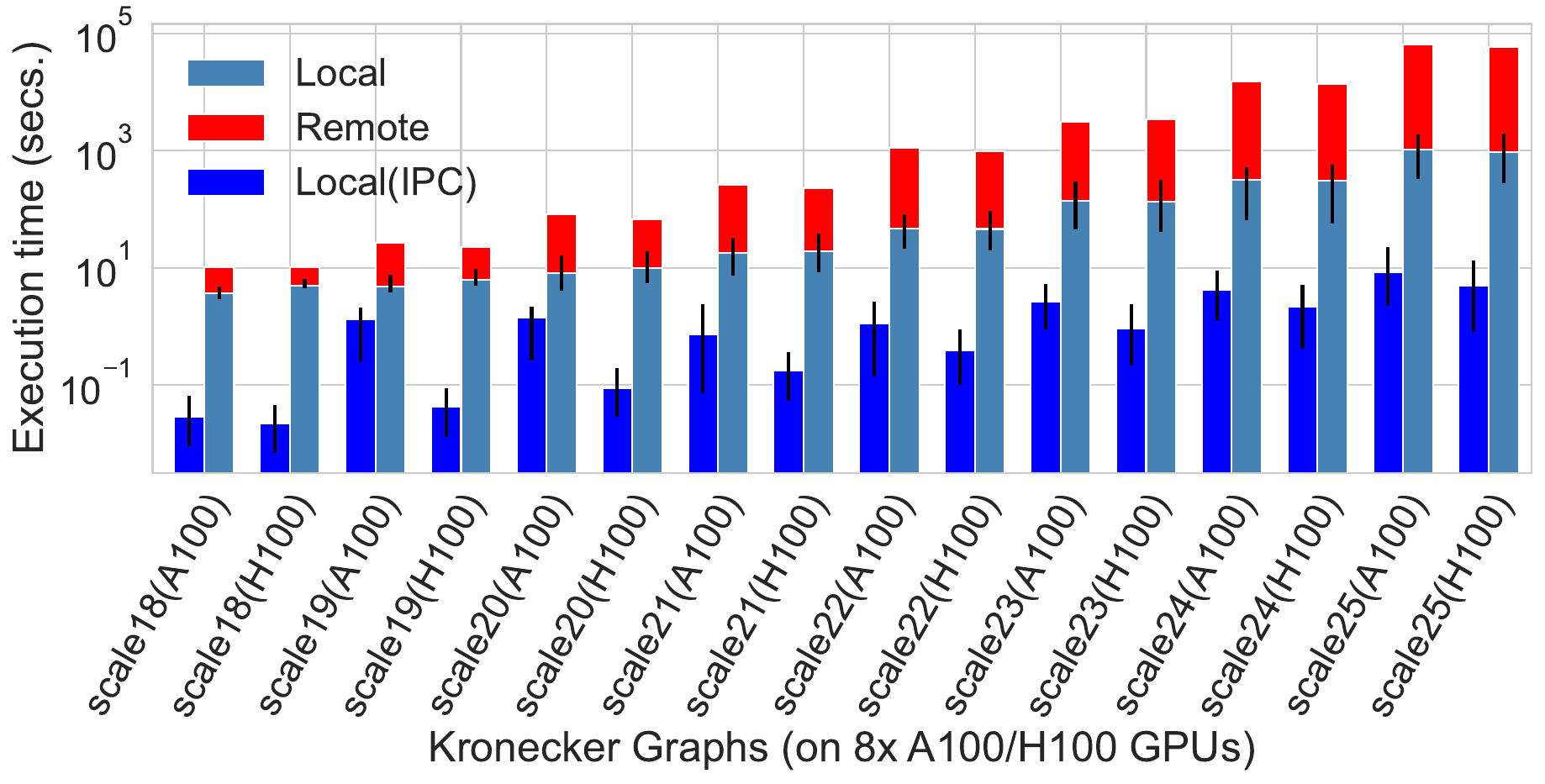}
\caption{\small Graph500 Kronecker graph (scales 20--25) evaluations on a single node: CPU vs. 8$\times$ A100(odd clusters)\slash H100 (even clusters).}
\label{fig:g500-cpu-dgx-a100-h100} 
\end{figure}

\begin{figure}[!ht]
\centering
\includegraphics[width=0.9\linewidth]{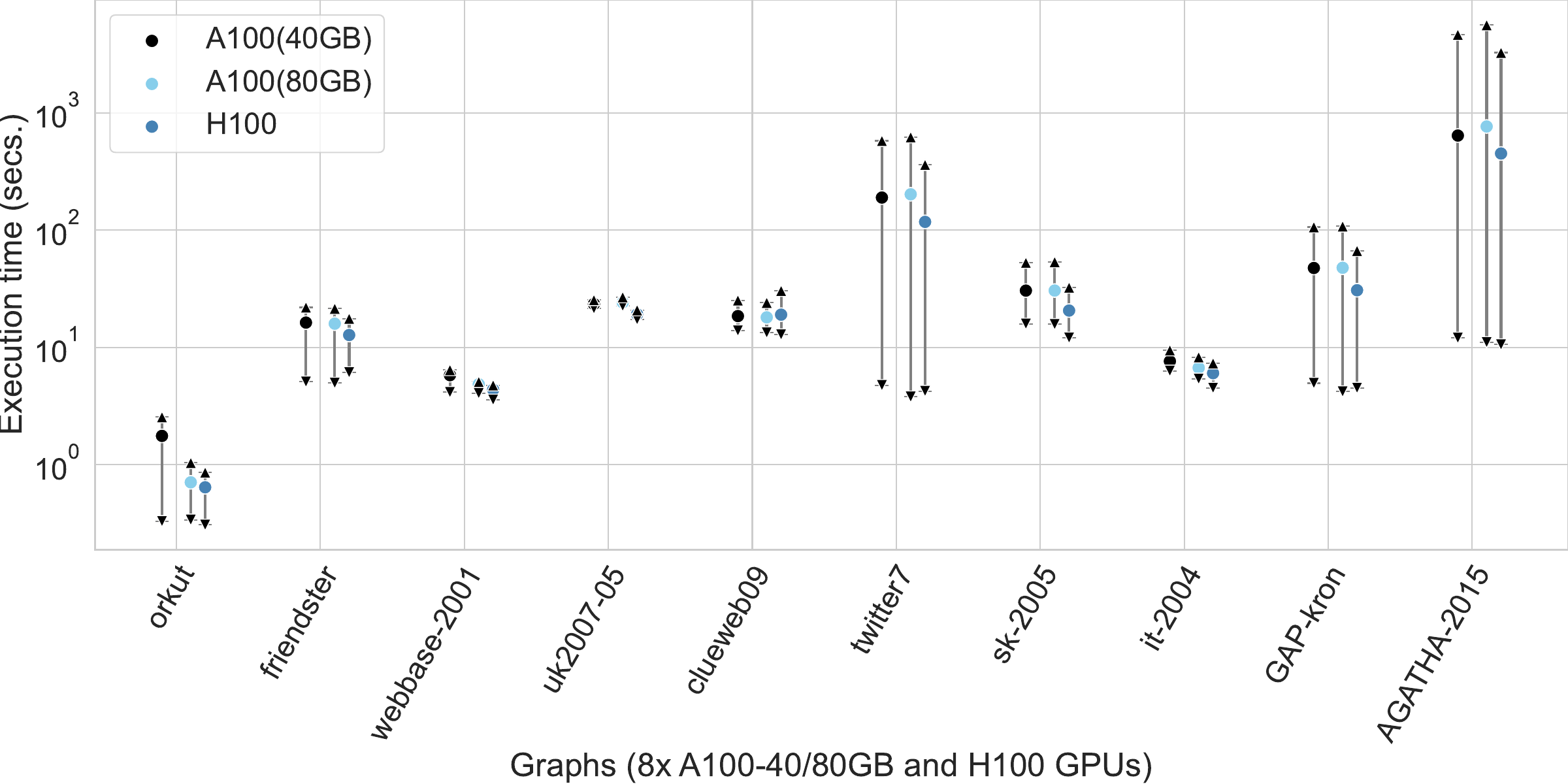}
\caption{\small Large graphs on 8$\times$ A100 \slash H100 DGX; \emph{dots}: average and \emph{arrows}: min--max  execution times (secs.).}
\label{fig:perf-dgx-a100-h100} 
\end{figure}

Contemporary dense-GPU systems such as NVIDIA DGX servers allow replacing remote communication entirely with IPC (i.e., ``zero copy'') over faster GPU interconnect fabric such as NVLink~\cite{foley2017ultra}. Our IPC-only variant is implemented as the masked kernel discussed in \S\ref{sssec:method-masked-edge} (reported times include host-side edge-masking and copies), and we evaluate on three DGX platforms with two GPU architectures---``Ampere'' (A100 with 40GB HBM and A100 with 80GB HBM2e) and ``Hopper'' (H100 with 40GB HBM3), see \S\ref{ssec:eval-setup-platforms}. First, we compare the IPC and MPI variants on A100 and H100 based DGX platforms using irregular scale-free graphs (\emph{scale}18--25, see Table~\ref{tab:datasets}: both MPI and IPC variant uses 8-way partitioned graphs) to keep MPI communication and load imbalances at minimum (scale 24--25 graphs exhibit major imbalances at scale). As shown in Fig.~\ref{fig:g500-cpu-dgx-a100-h100}, we observe the IPC variant to be multiple orders of magnitude faster, with 2$\times$ better performance on H100 platforms.
We also evaluated massive real-world graphs using IPC in Fig.~\ref{fig:perf-dgx-a100-h100}; many depict major startup issues for the MPI variants (requires more nodes and major communication overhead due to high intermediate data). The impact of faster NVLink (4th generation on DGX H100) lends to 20--30\% better performance overall.
Overall, pure-IPC dominates performance because it bypasses host-initiated buffering and network communication entirely, but local counting always benefits from increasing the \#GPUs, as shown in Table~\ref{tab:mpi-vs-ipc}. 
\begin{table}[!ht]
\centering
\caption{\small Comparing execution times (rounded, secs.) of MPI-IPC (with \textbf{L}ocal\slash \textbf{R}emote) on Polaris vs. pure-IPC (on DGX A100-40GB).}\label{tab:mpi-vs-ipc}
{\scriptsize
\begin{tabular}{|c|c||c|}
\hline
\textbf{Graphs} & \textbf{MPI-IPC (L + R), \#GPUs} & \textbf{IPC on 8$\times$ GPUs} \\ \hline \hline
friendster         & 11 (3(L) + 8(R)), 1024             & 16                               \\ \hline
webbase2001        & 29 (4(L) + 25(R)), 1024            & 6                                \\ \hline
uk2007-05           & 316 (4(L) + 312(R)), 512          & 23                               \\ \hline
clueweb09           & 17 (11(L) + 6(R)), 128            & 18                               \\ \hline
scale21             & 18 (1(L) + 17(R)), 512            & 1                                \\ \hline
scale22            & 45 (1(L) + 44(R)), 1024            & 1                                \\ \hline
scale23            & 105 (1(L) + 104(R)), 1024          & 3                                \\ \hline
\end{tabular}
}
\end{table}

\subsection{Workload Characteristics}\label{sec:profiling}
In distributed-memory, it is apparent that network communication is the bottleneck. However, when explicit data movement is supplanted by zero-copy IPC on dense-GPU systems (assuming saturation of the instruction pipelines), stalls emerge due to inherent irregularity of the workload leading to unpredictable under-utilization and owing to differences across the device generations. We discuss both of these aspects in forthcoming \S\ref{ssec:tail-effect} and \S\ref{sec:h100-vs-a100}, deliberately considering the scale20 input (see Table~\ref{tab:datasets}) due to its irregularity (standard deviation of edge distribution on 8$\times$ GPUs greater than 1000). 

\subsubsection{Tail effect}\label{ssec:tail-effect}
\begin{figure}[!ht]
\centering
\includegraphics[width=\linewidth]{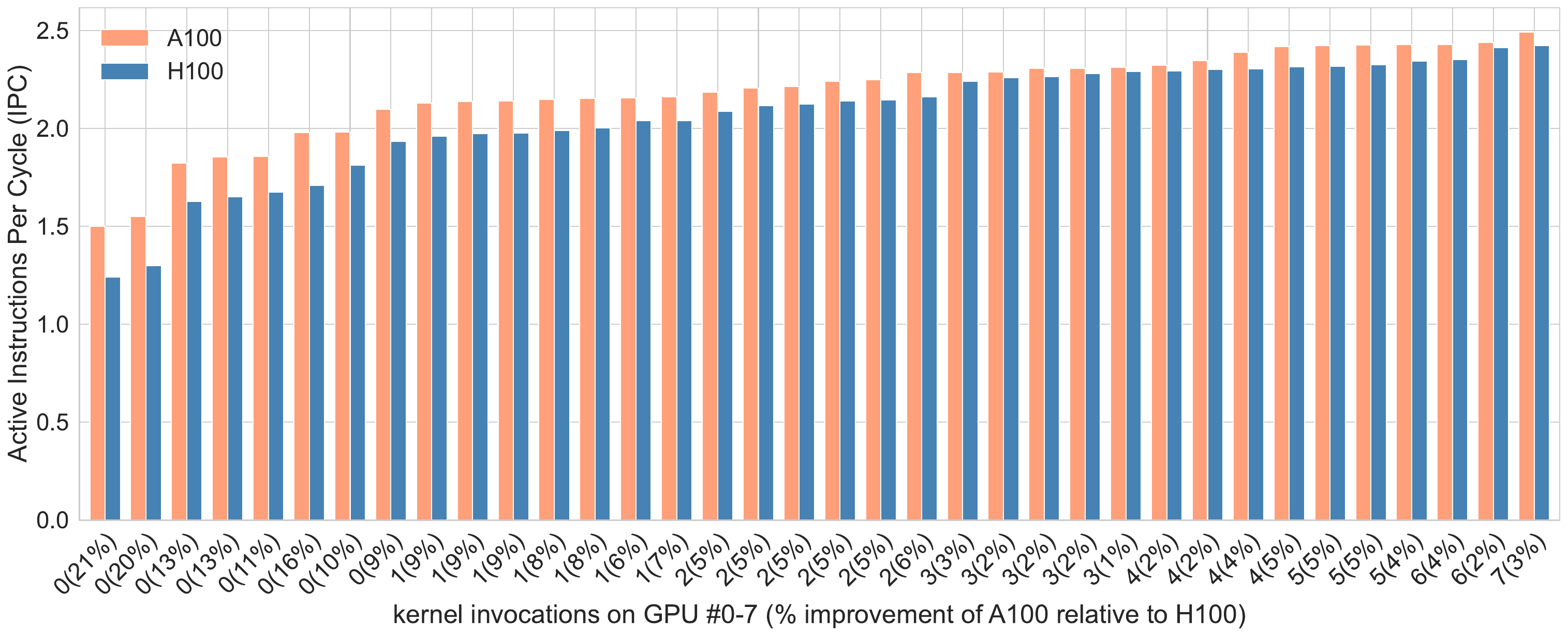}
\caption{\small Instructions Per Cycle (IPC) on A100 and H100.}
\label{fig:ipc_a100_h100_s20} 
\end{figure}

Despite zero FLOPs, this is a compute bound workload as evident from average Instructions Per Cycle (IPC) greater than 2 (see Fig.~\ref{fig:ipc_a100_h100_s20}, which reports IPC\slash kernel across GPUs) and high utilization of the ALU pipeline (over 75\% due to predominance of integer and logic operations). Irregular memory access patterns inherent in the triangle counting workload affects the readiness of active warps for issuing instructions, which has a second degree effect in limiting instruction throughput and achieved occupancy. Hardware supports up to 16 warps\slash scheduler (comprising of one transaction per cycle), and although we allocate over 70\% of the maximum warps possible (12\slash 16), only 2--5 warps could be materialized per cycle due to diverse execution durations per warp and associated data dependencies, leading to each warp stalling for a non-trivial \#cycles. 

Warp-level stalls and grid-stride pattern on irregular grids can ultimately contribute to the \emph{tail effect}, where nontrivial \#partial thread blocks can account for performance degradation. Partial thread-blocks originally arise from the variations in the remote edges post segregation (see \S\ref{sssec:method-ipc}) effecting kernel launches on limited blocks\slash grid, which is a byproduct of the structural irregularities of general graphs. For e.g., scale20 launch (on a grid of about 850--2000 blocks\slash kernel) resulted in 1--3 full waves (batch of thread blocks that can be executed concurrently on SMs) and a partial wave of 3--635 thread blocks across GPUs, which accounted for 20--35\% of the total execution time. This load imbalance leads to an achieved occupancy of about 60\%, the theoretical occupancy being 75\% for A100\slash H100 (estimated from block sizes and number of registers\slash shared memory). Controlling register allocation can mitigate the partial waves and improve occupancy (thereby tail effect) for a specific case (usually for smaller inputs), but it is highly nontrivial to balance parallelism and eventual register spillage considering arbitrary graphs across several concurrent kernel invocations on GPUs.

\subsubsection{A100 vs. H100}\label{sec:h100-vs-a100}
\begin{figure}[!ht]
\centering
\includegraphics[width=\linewidth]{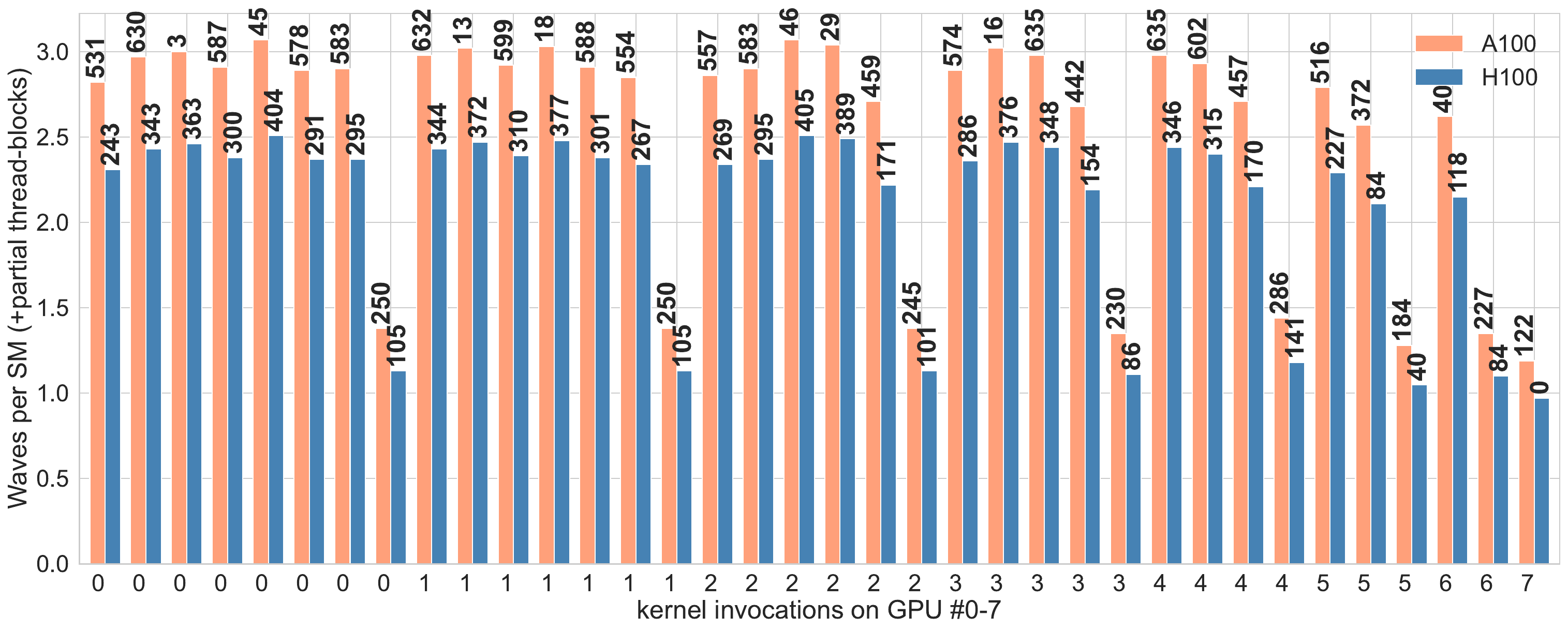}
\caption{\small Waves\slash SM and \#partial thread-blocks between A100 and H100; limited partial waves keep tail effect in check.}
\label{fig:waves_sm_ipc_s20} 
\end{figure}

\begin{figure}[!ht]
\centering
\includegraphics[width=\linewidth]{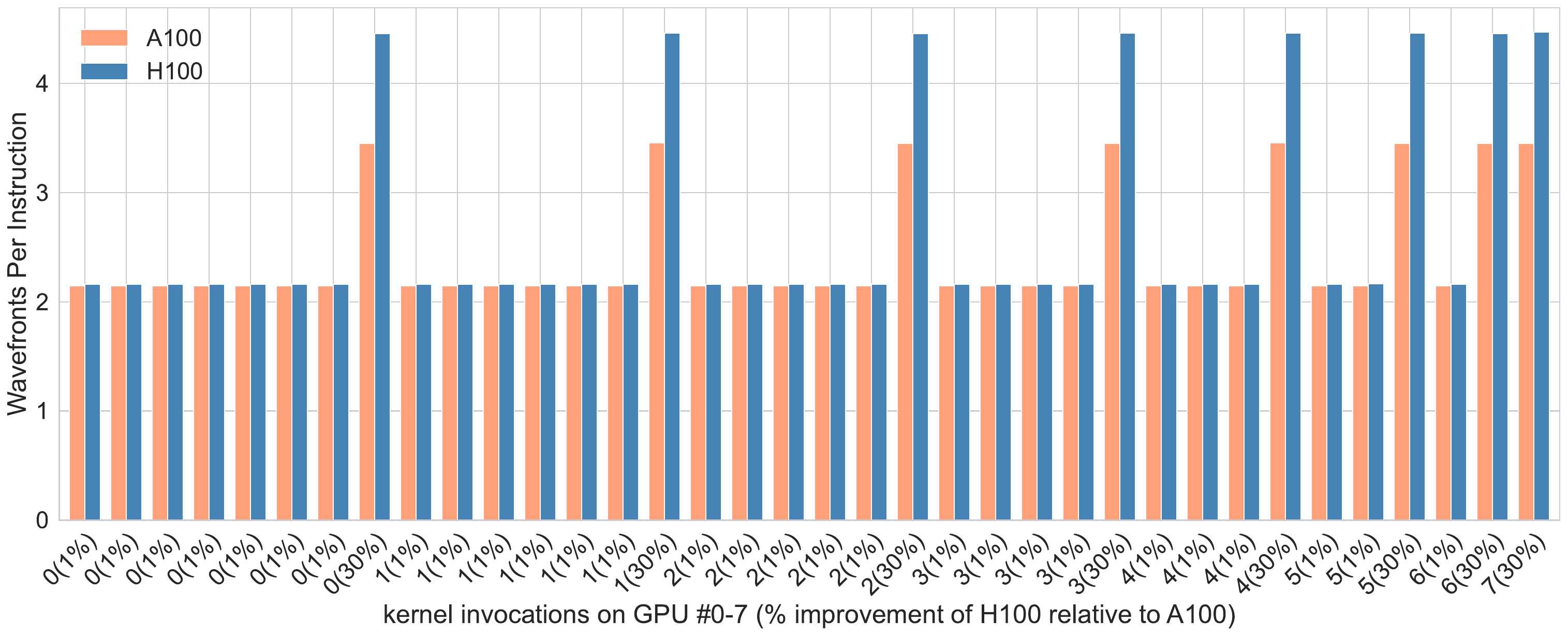}
\caption{\small \#Shared memory load\slash store wavefronts\slash cycle L1 cache operations between A100 and H100 (\emph{higher} is \emph{better}).}
\label{fig:waves_inst_ipc_s20} 
\end{figure}

\begin{figure}[!ht]
\centering
\includegraphics[width=\linewidth]{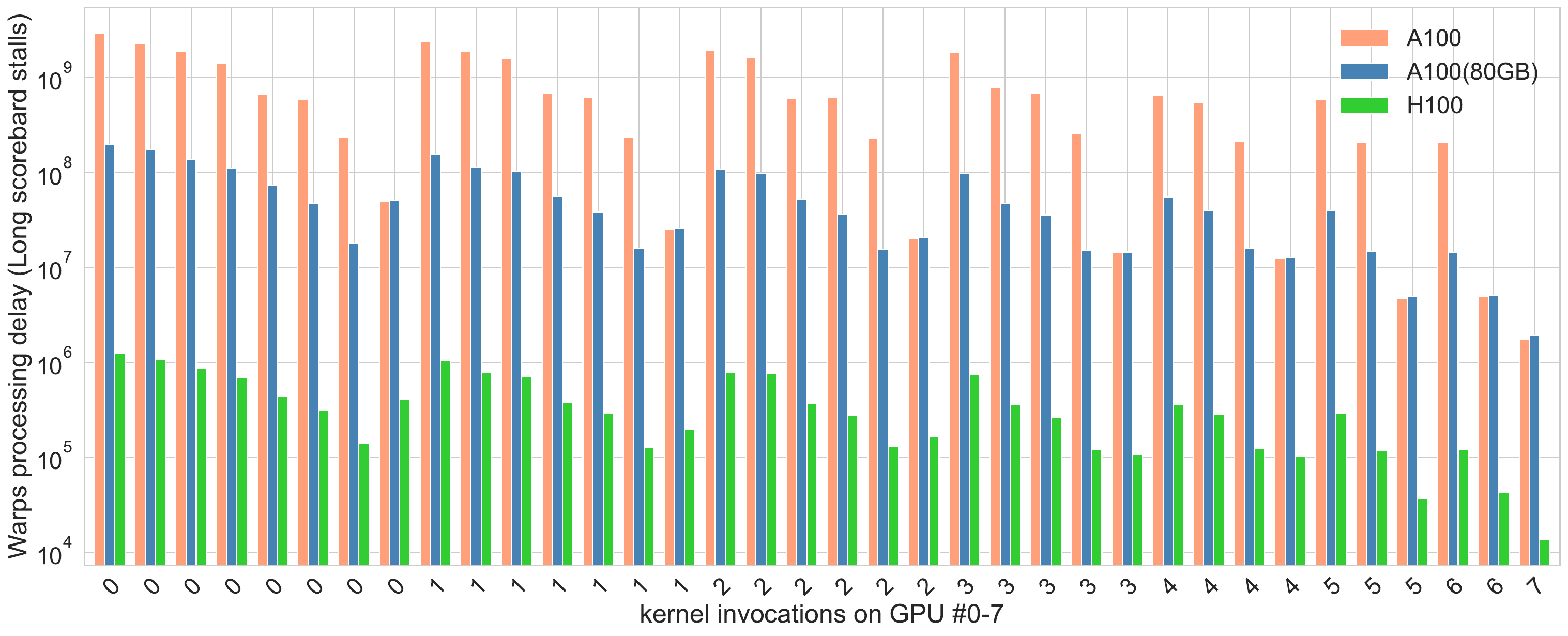}
\caption{\small \# Avg. instances of warp stalls between A100, A100(80GB) and H100 (\emph{lower} is \emph{better}).}
\label{fig:long_scoreb_ipc_s20} 
\end{figure}

A100 depicts up to 10\% better IPC as shown in Fig.~\ref{fig:ipc_a100_h100_s20}, since uncoalesced accesses (about 5--10\% of L2 sectors for this workload) and disparities in eligible warps can suffer from relatively higher latencies\slash stalls in H100 due to hierarchical L2 design and consistent integer\slash ALU pipeline rates between A100 and H100, despite 30\% higher clock speed and wider execution pipelines in H100. Despite the counterintuitive IPC, H100 runs consistently exhibit 2$\times$ better performance (see Fig.~\ref{fig:perf-dgx-a100-h100}) as compared to A100. Although H100 supports bypassing cache via asynchronous global-shared memory copy routines and dedicated inter-SM network~\cite{choquette2023nvidia}, maintaining a fixed-size shared memory array per thread-tile\slash warp (to store partial distance-2 neighborhoods, see Line 24 of Listing~\ref{lst:masked-edge-kernel}) increases overhead since prospective global-shared memory copies and irregular data exchange across SMs cannot be amortized by enhancing reuse in this case. The primary reason behind the performance disparity is due to fewer partial unassigned thread block slots on SMs because H100's per-SM shared memory capacity is about 40\% larger than A100, allowing the compaction of thread-blocks without hitting shared memory resource limits, and thereby reducing the leftover blocks can help regulate the tail effect. Fig.~\ref{fig:waves_sm_ipc_s20} depicts 2--4$\times$ reduction in the partial blocks across kernel invocations on GPUs, while demonstrating about 20\% better compaction.
The performance windfall from the reduction in partial blocks can be attributed to kernel launches resulting in a single full wave (usually the last kernel invoked on a GPU with the edges local to a GPU). For e.g., on GPU\#7, the only kernel launch resulted in 769 thread blocks, materializing into a single full wave and 122 partial thread blocks on A100, whereas there were zero leftovers on H100. Consequently, we observe up to 30\% better shared memory throughput for the kernels on the least \#partial thread-blocks (in terms of wavefronts\slash instruction, which measures the maximum \#loads\slash stores handled by the L1 data pipeline) in H100 (since shared memory is 32-way on A100\slash H100, 32 is the theoretical maximum), as shown in Fig.~\ref{fig:waves_inst_ipc_s20}.  
Finally, we also measure \#instances on average when warps were stalled on data from L1 cache (e.g., pertaining to long-latency data accesses) for A100 and H100 platforms. Due to better compaction of the thread-blocks across SM resources in H100 (additionally 4th generation NVLink offers 50\% higher bandwidth~\cite{choquette2023nvidia}), we observe two orders of magnitude reduction in the number of stalled warps, as shown in Fig.~\ref{fig:long_scoreb_ipc_s20}.
Since compaction of the thread-blocks in H100 improved throughput for kernels with the least leftover partial thread-blocks (Fig.~\ref{fig:waves_sm_ipc_s20}), we investigate potential optimizations next to further squeeze them into a single wave, eradicating any leftovers.

\subsubsection{Further optimizations on H100}\label{sec:h100-imp}
\begin{figure*}[!ht]
     \centering
     \begin{subfigure}[b]{0.27\textwidth}
         \centering
         \includegraphics[width=\textwidth]{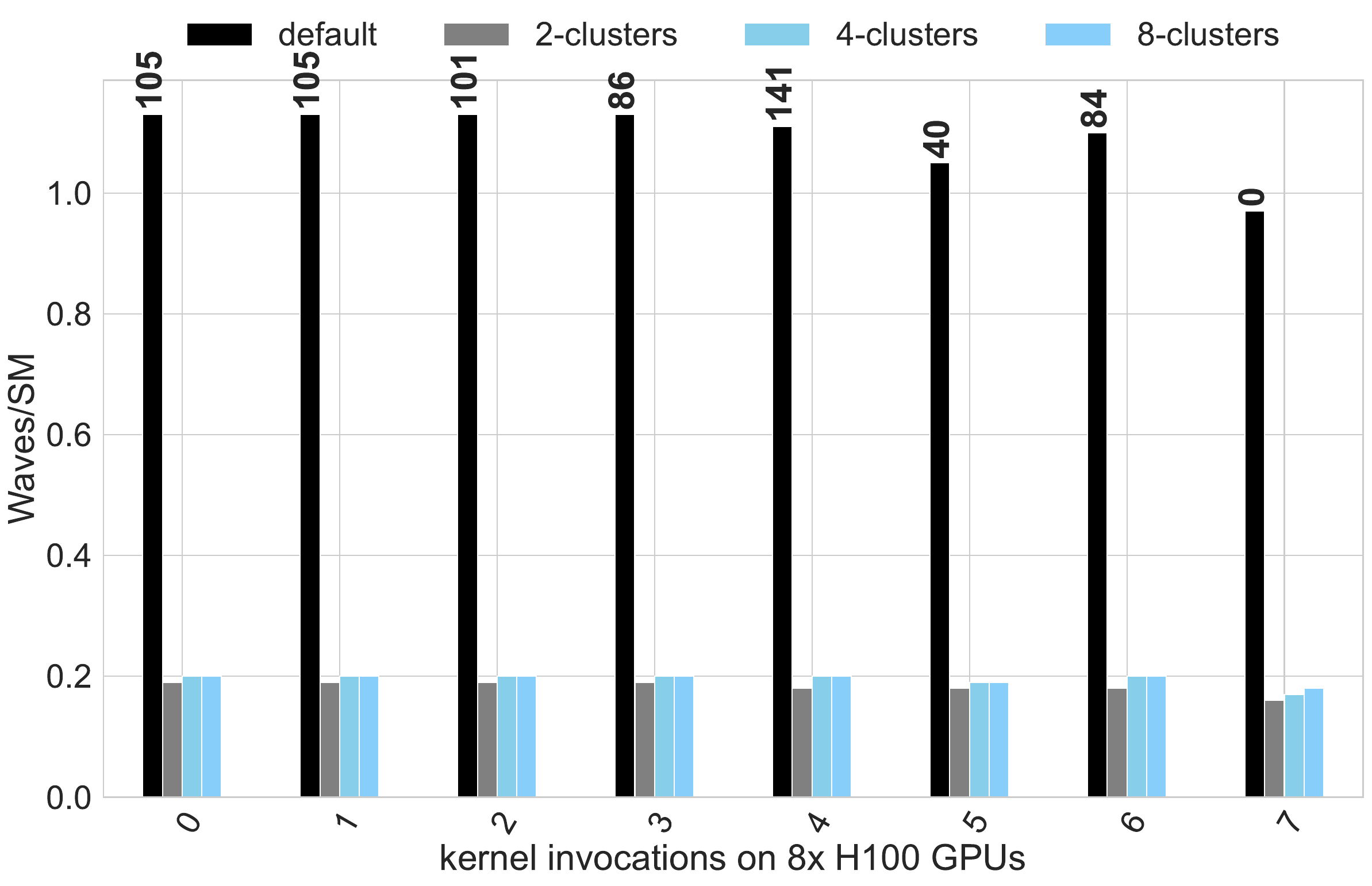}
         \caption{\small Waves\slash SM (0-partial w clusters).}
         \label{fig:waves_sm_clusters}
     \end{subfigure}
               \hfill
     \begin{subfigure}[b]{0.32\textwidth}
         \centering
         \includegraphics[width=\textwidth]{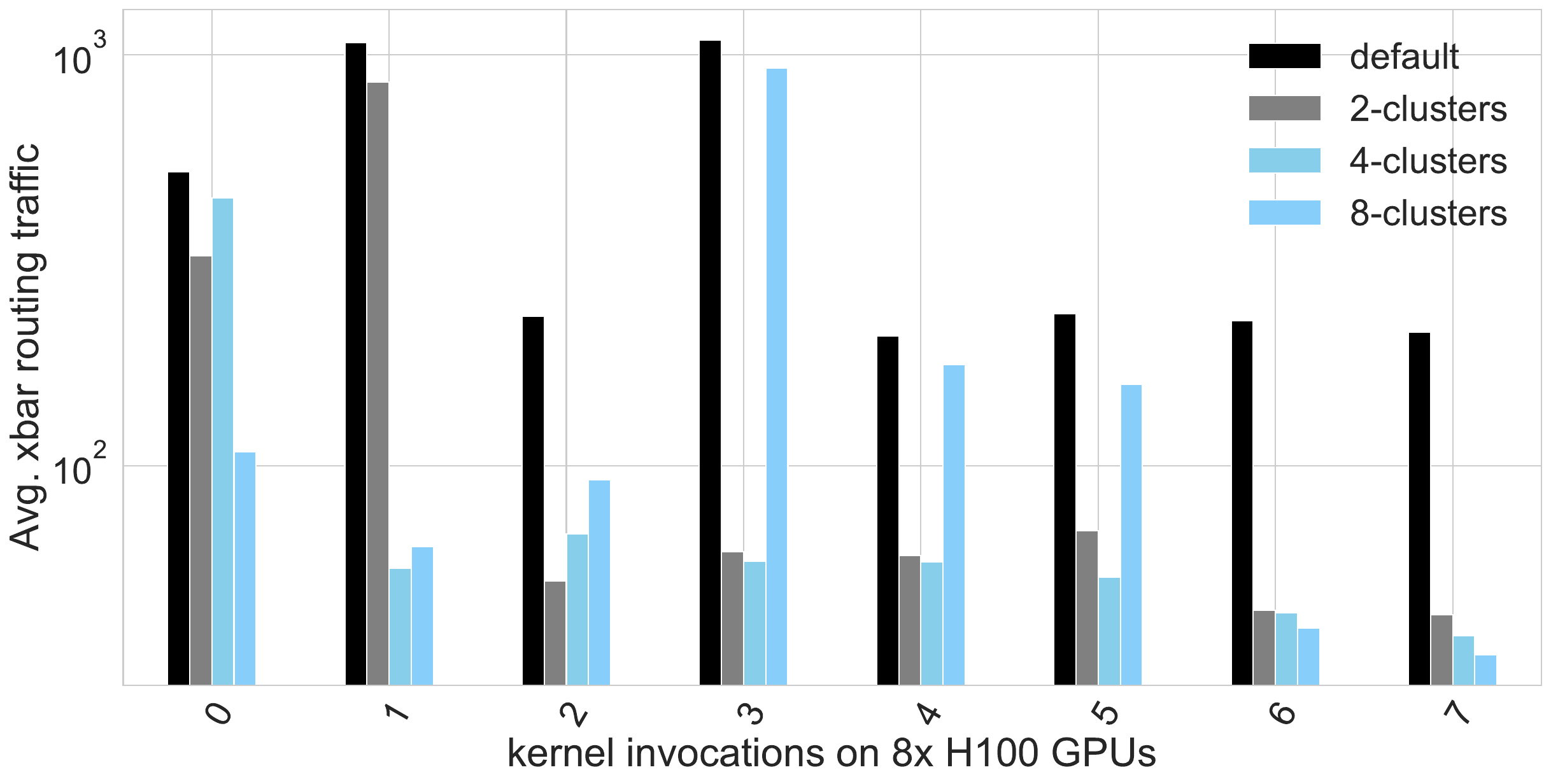}
         \caption{\small Crossbar bypass with clusters.}
         \label{fig:xbar_bypass}
     \end{subfigure}
     \hfill
     \begin{subfigure}[b]{0.32\textwidth}
         \centering
         \includegraphics[width=\textwidth]{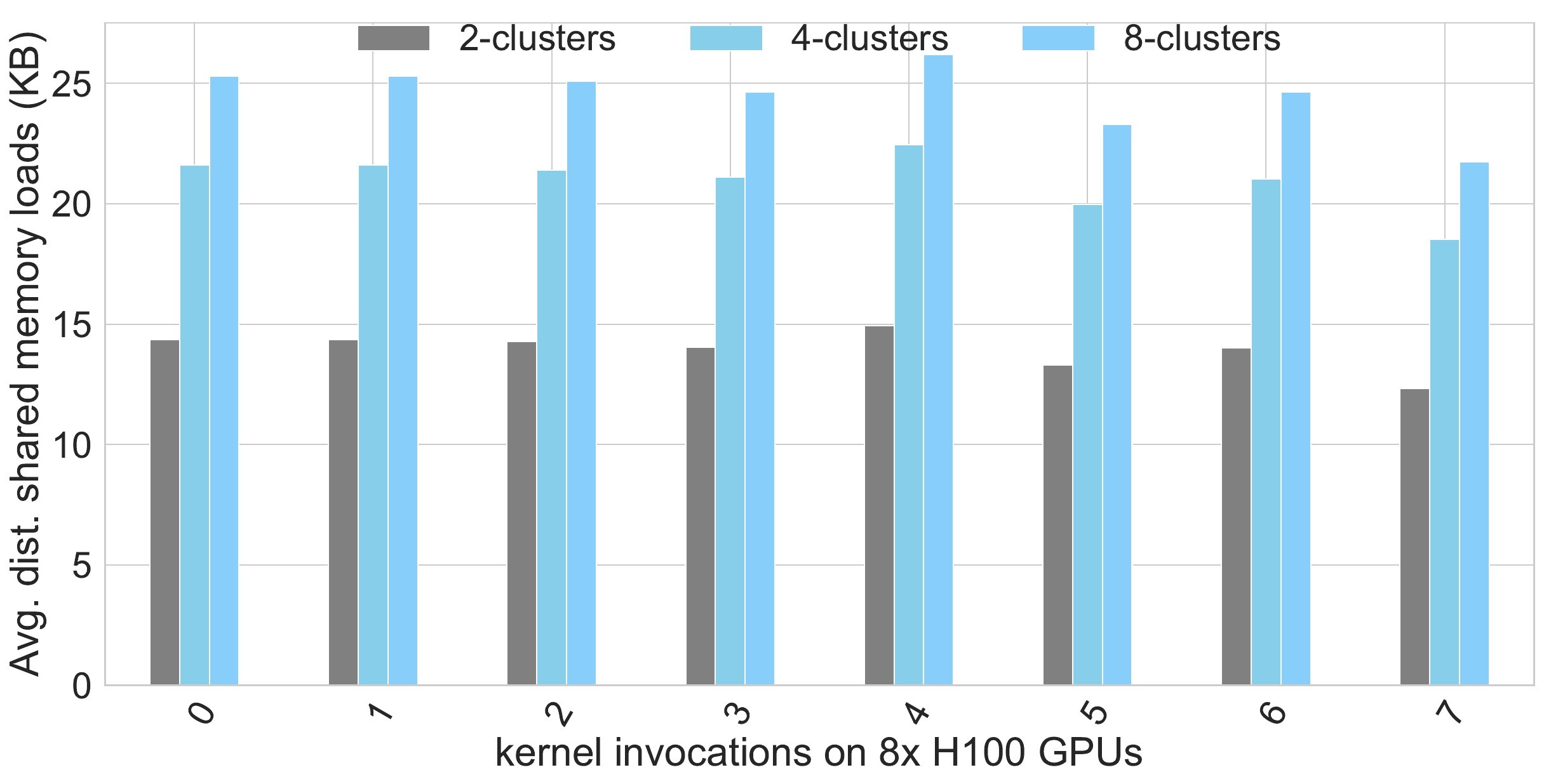}
         \caption{\small DSMEM volumes across clusters.}
         \label{fig:dsmem_bytes}
     \end{subfigure}
        \caption{\small Effect of thread-block compaction on H100 using clusters for node-local enumerations.}
        \label{fig:h100_clusters_opt}
\end{figure*}
H100 is organized into 8 Graphics Processing Clusters (GPCs), allowing applications to exploit inter-SM locality for reducing the round-trips to global HBM~\cite{choquette2023nvidia}. Instead of scheduling individual thread-blocks at a time, thread-blocks within a \emph{cluster} are scheduled simultaneously within a GPC. This coarser scheduling granularity allows further compaction of the launch grid, especially for kernels already using the least \#thread-blocks, thereby fitting into fewer hardware \#waves (in this case, a single wave). We modify Lines 33--37 of Listings~\ref{lst:masked-edge-kernel} (padding grid as necessary to be a multiple of \#clusters) to facilitate device-wide cluster-level reduction for arbitrary cluster sizes.
\begin{figure}[!ht]
\centering
\includegraphics[width=\linewidth]{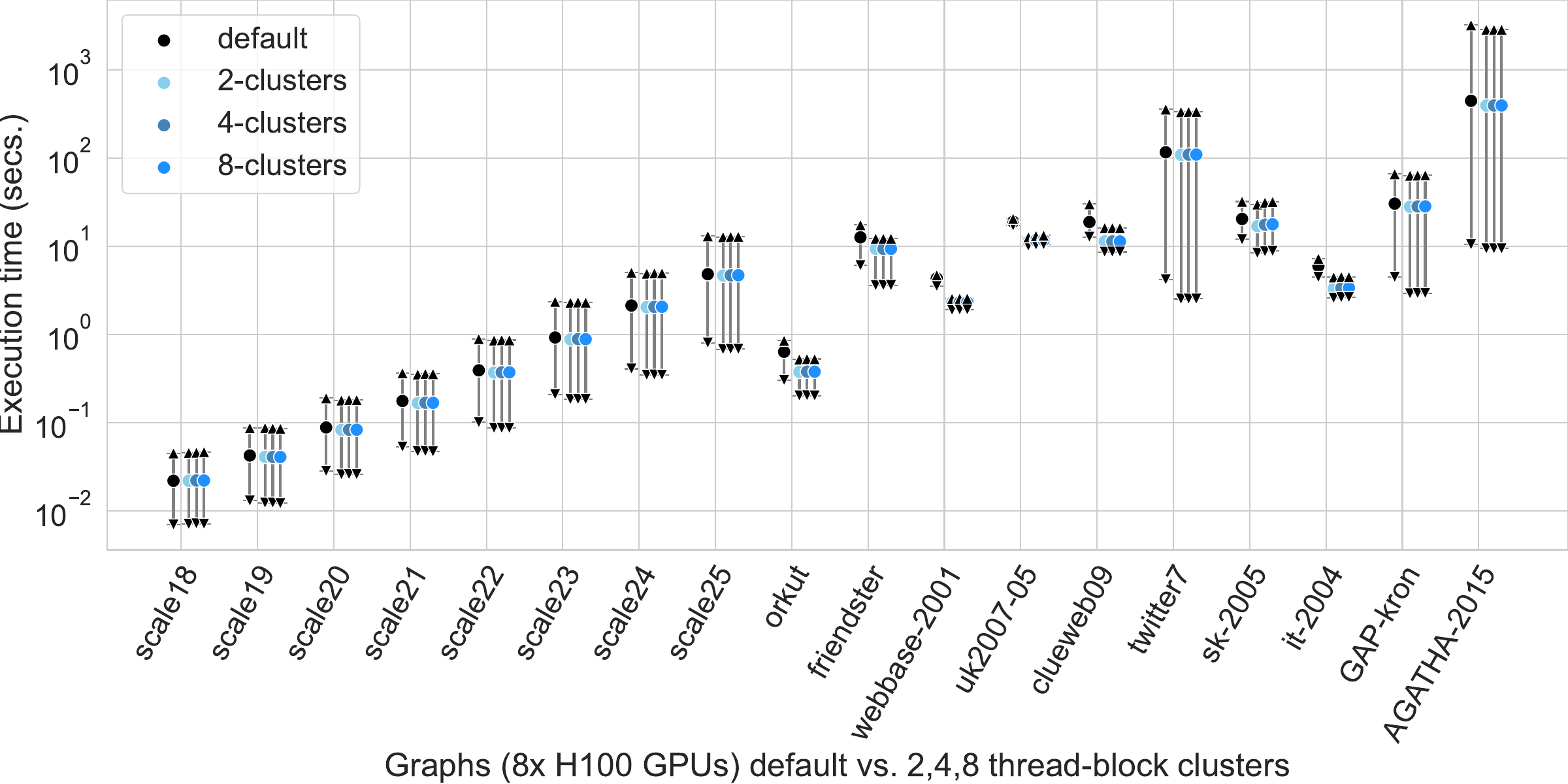}
\caption{\small Default vs. clusters on H100 (min--max performance range).}
\label{fig:clusters_h100} 
\end{figure}
For graphs with relatively high node-local enumerations (see Fig.~\ref{fig:local-accesses-perlmutter}\slash ~\ref{fig:local-accesses-polaris}), using 2-clusters, we observe up to 40\% improvement in the execution times, whereas for the scale-free graphs with power law degree distribution (i.e., scale 18--25, see Table~\ref{tab:datasets}), the improvements are around 12\% due to relatively low node-local accesses, as depicted in Fig.~\ref{fig:clusters_h100}. Increasing the \#clusters beyond 2 amplifies the amount of partial waves due to scheduling constraints (issuing \#thread-blocks\slash cluster on available resources concurrently), slightly exacerbating the tail effect (e.g., in Fig.~\ref{fig:waves_sm_clusters} we observe 10\% increase in the partial waves between successive cluster configurations, increasing the gap between issuing consecutive instructions). Due to this structural imbalance and enhanced SM crossbar traffic, we speculate that increasing clusters may lead to contention owing to poll-based synchronizations between thread-block clusters while performing reduction. Consequently, Distributed-shared Memory (DSMEM) volumes are 2$\times$ between 2 and 4\slash 8 clusters, as shown in Fig.~\ref{fig:dsmem_bytes}.

Cache lines (4 sectors $\times$ 32 bytes = 128 bytes) are associated with tags, and data presence is determined through tag and sector-data lookups; when a request bypasses L2 cache, it skips tag lookup. Blocks within the same cluster can directly access each other's SMs, bypassing regional cache processing pipelines. Fig.~\ref{fig:xbar_bypass} lists the average \#memory requests routed from SM crossbars to the regional L2 cache\slash memory subsystem for default and cluster configurations, depicting about 2--5$\times$ reduction in crossbar routing. Without clusters, thread blocks are spread across the GPU (less compaction, more partial \#waves: see Fig.~\ref{fig:waves_sm_clusters}), streaming requests through the crossbar to specific memory region controllers. Clusters improve the proximity of the thread blocks, localizing inter-block traffic within a cluster, eliminating excess global memory requests routing through the crossbar.

\section{Comparisons with GPU graph frameworks}\label{sec:related}
\begin{figure}[!ht]
\centering
\includegraphics[width=\linewidth]{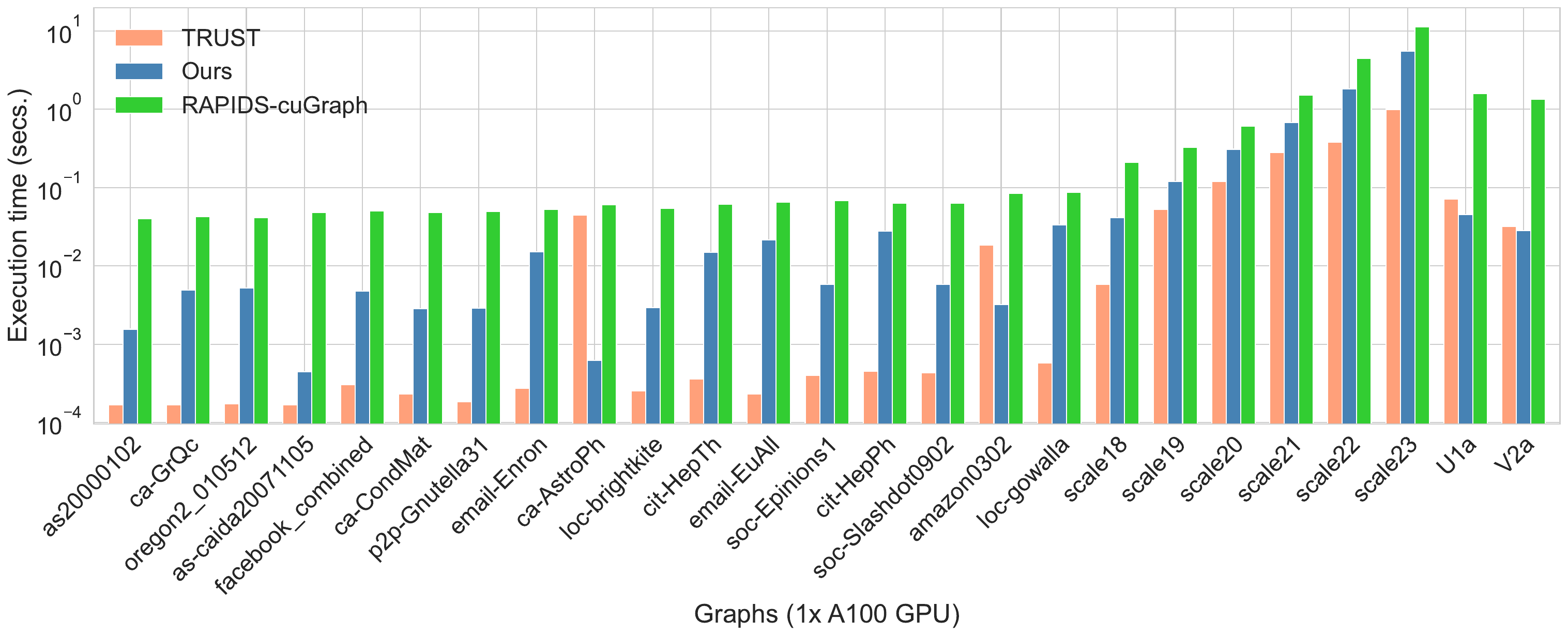}
\caption{\small Single GPU (A100) execution time comparisons between TRUST, NVIDIA cuGraph and our implementation.}
\label{fig:single-gpu-perf-all} 
\end{figure}

\begin{figure}[!ht]
\centering
\includegraphics[width=0.9\linewidth]{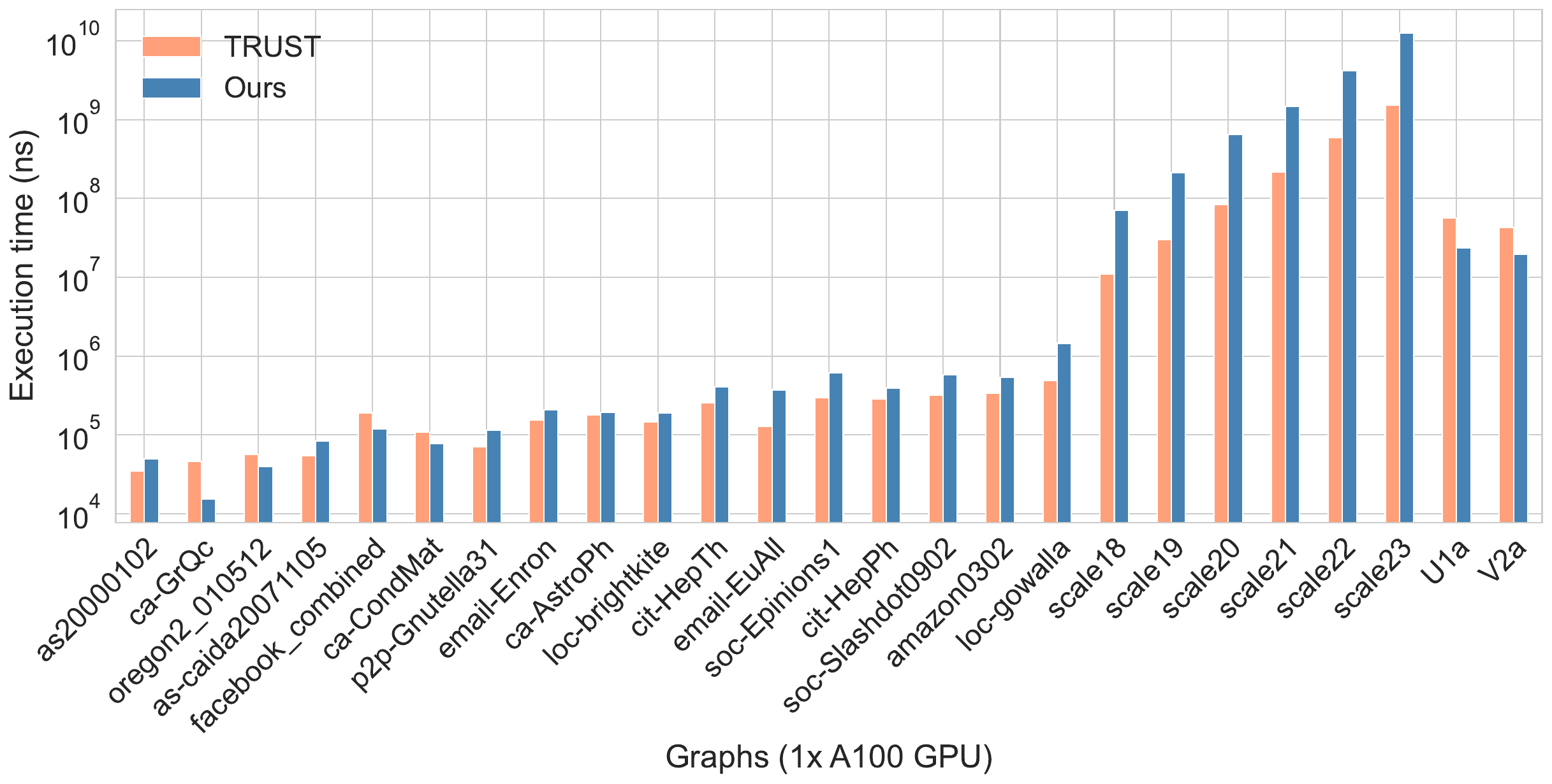}
\caption{\small Single GPU (A100) kernel execution time (in nsecs.) comparisons between TRUST and our implementation.}
\label{fig:single-gpu-trust-kernel} 
\end{figure}

\paragraph*{Single GPU evaluations} We assess performance against established state-of-the-arts: NVIDIA RAPIDS\regtm{} cuGraph~\cite{fender2022rapids} and TRUST~\cite{pandey2021trust}. NVIDIA cuGraph makes heavy use of C++ template metaprogramming in implementing edge-intersection based triangle counting, whereas TRUST incorporates a vertex-centric hashing-based method to augment the edge-intersection method (maintaining hash table at the expense of memory\slash compute). Theoretically, hashing based approaches are faster ($\mathcal{O}(1)$ vs. $\mathcal{O}(\log N$) for search), but edge-based search strategy is more flexible as it is directly applicable to the underlying graph data structure. Our implementation is not particularly optimized for single-GPU, but we observe 2--100$\times$ improvement over state-of-the-art vendor optimized cuGraph for a variety of graphs in Fig.~\ref{fig:single-gpu-perf-all} (inputs in Table~\ref{tab:datasets}). Further, cuGraph depicts major device-device data transfer overheads (nil for us\slash TRUST, whereas at least an order-of-magnitude higher than cuGraph host-device) due to the intermediate copies owing to C++ generic function template based implementation. TRUST end-to-end improvements can also be attributed to adopting 32-bit \emph{int} CSR vertex\slash index data types, compared to 64-bit graph CSR for us\slash cuGraph (doubling memory transfer overheads at different levels), and TRUST excluding the relevant memory allocation\slash copy times from the reported final time (profiles indicate competitive performance to TRUST for most of the degree-balanced graphs). 
Upon profiling TRUST and our implementation to compare the respective kernel execution times, despite the overheads of search-based approach (particularly inimical for heavy-tailed power law graphs such as \emph{scale18--23}) as shown in Fig.~\ref{fig:single-gpu-trust-kernel}) and 32 vs. 64-bit datatype disparities, we observe competitive performance to TRUST for most of the degree-balanced graphs.  
\begin{figure}[!ht]
\centering
\includegraphics[width=\linewidth]{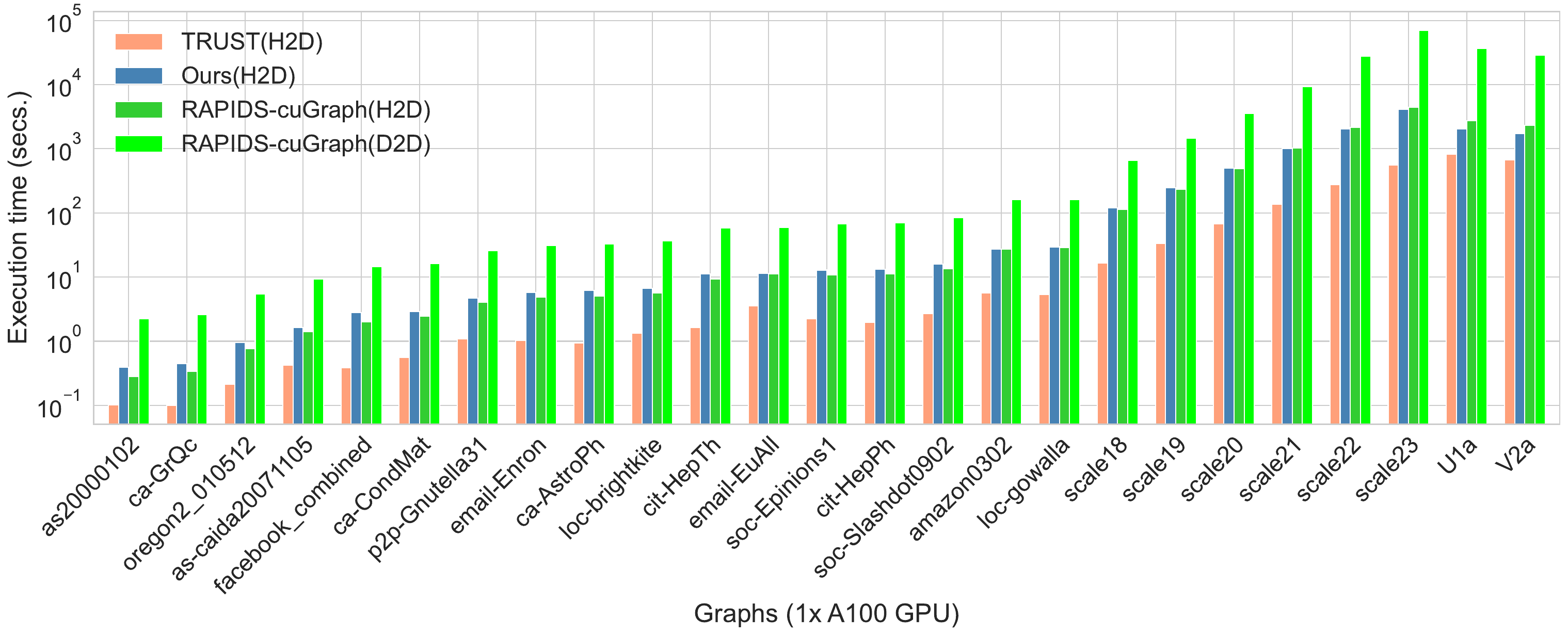}
\caption{\small Single GPU (A100) host-device (H2D) and on-device (D2D) memory transfer (in MBs) comparisons between TRUST, NVIDIA cuGraph and our implementation.}
\label{fig:single-gpu-memcpy} 
\end{figure}
We also investigate host-device and device-device data transfers to analyze the single-GPU evaluations. 
TRUST exhibits the least (host-device) memory copies due to 32-bit \emph{int} CSR representation, whereas cuGraph exhibits nontrivial device-device data transfer overheads (nil for us\slash TRUST, whereas at least an order-of-magnitude higher than cuGraph host-device) due to the intermediate copies owing to C++ generic function template based implementation, as depicted in Fig.~\ref{fig:single-gpu-memcpy}. Although the original TRUST paper~\cite{pandey2021trust} discusses distributed-memory multi-GPU evaluations via 2D-partitioning (i.e., edge replication), multi-GPU variant of TRUST presently crashes with a segmentation fault.
\paragraph*{Multi-GPU evaluations} 
Working multi-GPU triangle counting implementations are rare, we compare against GraphMiner\footnote{\url{https://github.com/chenxuhao/GraphMiner}}, a benchmarking suite for graph pattern matching. Graphminer includes OpenMP-based CPU and GPU implementations (single node, multi-GPU), building on SandLash~\cite{chen2021sandslash} and $G^2$Miner~\cite{chen2022efficient}, performing pattern-specific preprocessing and workload-balanced scheduling (even-split or round-robin with fixed chunksize of 1024) for intersection based triangle counting on GPUs. We perform multi-GPU comparisons on NVIDIA DGX platform (\S\ref{ssec:eval-setup-platforms}). 
\begin{figure}[!ht]
\centering
\includegraphics[width=0.9\linewidth]{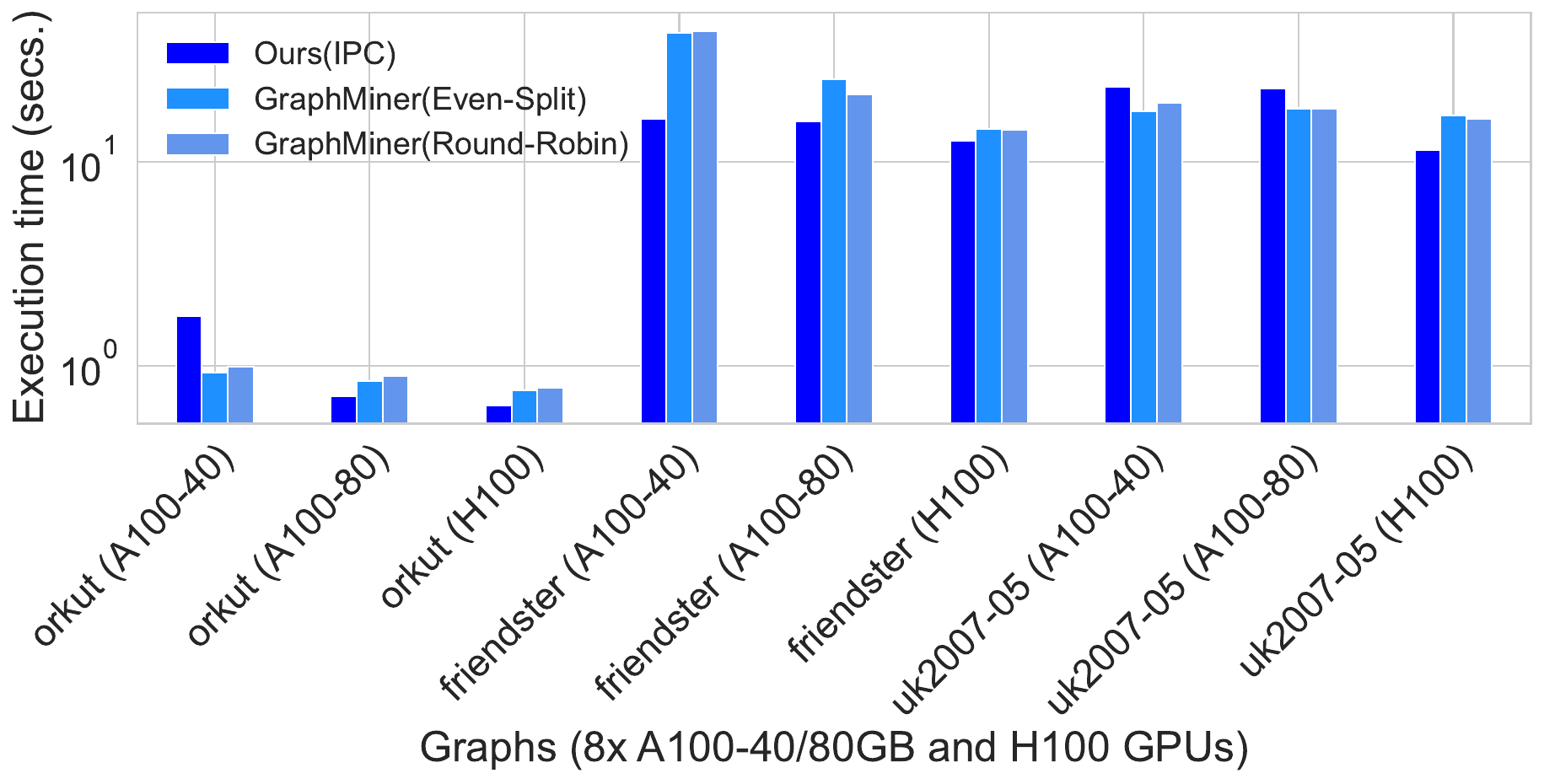}
\caption{\small GraphMiner vs. our baseline (IPC) implementation.}
\label{fig:dgx-graphminer} 
\end{figure}
As shown in Fig.~\ref{fig:dgx-graphminer}, GraphMiner is 20--40\% faster on 2/3 graphs than our baseline IPC implementation (i.e., cluster=1) on DGX A100 (40/80GB) platforms, but on H100 platforms, we exhibit 10-60\% better performance for all scenarios. In GraphMiner, DAG creation (schedule generation) took about 10--80\% of the total execution time, whereas the remaining time is divided among edge-list generation, device copy and kernel execution. Further, our implementation is poised to improve by greater than 2$\times$ between GPU architectures with enhanced memory bandwidths.

\section{Concluding Remarks}\label{sec:summary}
Data movement costs for irregular graph workloads are further accentuated on GPU platforms since processing is relatively faster (than CPUs, even with host-device transfers). However, there are options to tweak concurrency and data movement by leveraging hybrid CPU\slash GPU and ``zero copy'' mechanisms. This paper discusses the impact of these options by deconstructing the triangle counting problem into the fundamental unit of parallel edge enumeration tasks.

Overall, IPC demonstrates superior performance compared to the MPI-GPU variants due to zero intermediate data. However, distributed-memory variants can conveniently adapt to larger and sparser inputs, as long as the load imbalance induced by the graph structure can be alleviated. Comparisons with the related GPU approaches demonstrates that limiting host operations and intermediate copies is a path toward sustainable performance across GPU generations. 


\section*{Acknowledgments}
This research was supported by the Data-Model Convergence (DMC), a multi-disciplinary Pacific Northwest National Laboratory (PNNL) initiative that applies co-design to develop heterogeneous computing capabilities to tackle challenging scientific problems. PNNL is operated by Battelle Memorial Institute under Contract DE-AC06-76RL01830. This research also used resources of National Energy Research Scientific Computing Center (NERSC) and Argonne Leadership Computing Facility (ALCF), Department of Energy Office of Science User Facilities.
\bibliographystyle{IEEEtran}
\bibliography{paper}

\end{document}